\documentclass[a4paper]{article}
\usepackage{lmodern}
\usepackage{amssymb,amsmath}
\usepackage{ifxetex,ifluatex}
\usepackage{fixltx2e} 
\ifnum 0\ifxetex 1\fi\ifluatex 1\fi=0 
  \usepackage[T1]{fontenc}
  \usepackage[utf8]{inputenc}
\else 
  \ifxetex
    \usepackage{mathspec}
  \else
    \usepackage{fontspec}
  \fi
  \defaultfontfeatures{Ligatures=TeX,Scale=MatchLowercase}
\fi
\IfFileExists{upquote.sty}{\usepackage{upquote}}{}
\IfFileExists{microtype.sty}{%
\usepackage{microtype}
\UseMicrotypeSet[protrusion]{basicmath} 
}{}
\usepackage[margin=1in]{geometry}
\usepackage{rotating,caption,booktabs}
\usepackage{hyperref,natbib}
\hypersetup{unicode=true,
            pdftitle={Rank-based approach for estimating correlations in mixed data},
            pdfauthor={Xiaoyun},
            pdfborder={0 0 0},
            breaklinks=true}
\usepackage{graphicx,grffile}

\makeatletter
\def\maxwidth{\ifdim\Gin@nat@width>\linewidth\linewidth\else\Gin@nat@width\fi}
\def\maxheight{\ifdim\Gin@nat@height>\textheight\textheight\else\Gin@nat@height\fi}
\makeatother
\setkeys{Gin}{width=\maxwidth,height=\maxheight,keepaspectratio}
\IfFileExists{parskip.sty}{%
\usepackage{parskip}
}{
\setlength{\parindent}{0pt}
\setlength{\parskip}{6pt plus 2pt minus 1pt}
}
\ifx\paragraph\undefined\else
\let\oldparagraph\paragraph
\renewcommand{\paragraph}[1]{\oldparagraph{#1}\mbox{}}
\fi
\ifx\subparagraph\undefined\else
\let\oldsubparagraph\subparagraph
\renewcommand{\subparagraph}[1]{\oldsubparagraph{#1}\mbox{}}
\fi

\let\rmarkdownfootnote\footnote%
\def\footnote{\protect\rmarkdownfootnote}

\usepackage{titling}

  \title{\LARGE\bf Rank-based approach for estimating correlations in mixed ordinal data} 
 
   \author{Xiaoyun Quan,  James G. Booth\thanks{Professor Booth's research was partially supported
by an NSF grant NSF-DMS 1611893.}\; and Martin T. Wells\thanks{Professor Wells' research was partially supported by NSF-DMS 1611893 and NIH grant U19 AI111143.} \hspace{.2cm}\\
    Department of Biological Statistics and Computational Biology\\
 Department of Statistical Science, Cornell University, Ithaca NY, 14853, USA.}

 \date{August, 2018}

\usepackage{color}
\usepackage{bbm}
\usepackage{mathtools}
\usepackage{cancel}
\usepackage[utf8]{inputenc}
\usepackage[english]{babel}
\usepackage{amsthm}

\DeclareMathOperator*{\argmin}{arg\,min}
\usepackage[utf8]{inputenc}
\usepackage[english]{babel}

\begin{document}
\maketitle

\begin{abstract} \label{abstract}
High-dimensional mixed data as a combination of both continuous and ordinal variables are widely seen in many research areas such as genomic studies and survey data analysis. Estimating the underlying correlation among mixed data is hence crucial for further inferring dependence structure. We propose a semiparametric latent Gaussian copula model for this problem. We start with estimating the association among ternary-continuous mixed data via a rank-based approach and generalize the methodology to p-level-ordinal and continuous mixed data. Concentration rate of the estimator is also provided and proved. At last, we demonstrate the performance of the proposed estimator by extensive simulations and two case studies of real data examples of algorithmic risk score evaluation and cancer patients survival data. 
\end{abstract}

\noindent%
{\it Keywords:}  Algorithmic fairness, Gaussian copula model, graphical models, Kendall's $\tau$, latent variable models, mixed data, nonparanormal distribution, ordinal data, sparse modeling

\fontsize{11}{18} \selectfont


\section{1 Introduction}\label{introduction}

High-dimensional multilevel ordinal and continuous mixed data are now routinely collected in many research areas. For example, questionnaires and rating scales are commonly used to measure qualitative variables, such as attitudes and many behavioral, health-related variables and risk score measures and their relation to binary and continuous variables such as income, age, background history, or a phenotype. Another example is that of single nucleotide polymorphism (SNP) and expression data in genetics. Multilevel ordinal variables often arise as a result of discretizing latent continuous variables \citep{SkrondalR2007}. \cite{Fanetal2017} propose a generative latent Gaussian copula model for binary and mixed data, assuming the binary data are obtained by dichotomizing a continuous latent variable.  Multilevel ordinal data are common in survey data as well.    Estimating the associations between mixed data types is of great importance to gain insights about dependence between the variables, particularly for conditional dependencies and potential causal pathways. 


There are several classical rank-based methods for analyzing association among ordinal variables \citep{Agresti2010}. Specifically, those measures are all based on the numbers of concordant and discordant pairs of observations. A pair of observations, say $(X_i,Y_i)$ and $(X_{i'}, Y_{i'})$, is concordant if the subject that has a higher ranking on $X$ also has a higher ranking on $Y$, and on the other hand this pair is called discordant if the subject ranking higher on $X$ ranks lower on $Y$. Kendall's \textit{tau-a} ($\tau^a$) was first proposed by \citep{kendall1938} as a measure that quantifies the difference between proportions of concordant and discordant pairs among all pairs, which is essentially a correlation coefficient for sign scores. Later in 1945, a revised version called \textit{tau-b} ($\tau^b$) was introduced \citep{kendall1945} that took tied pairs into consideration. \citep{goodman} proposed the \textit{gamma} measure as the difference between proportions of concordant and discordant pairs among all concordant and discordant pairs. Other similar measures such as Somers' \textit{d} \citep{somers} also considers the difference between proportions of concordant and discordant pairs, just with a different base  as its denominator. Yet it remains an open question to measure the association between ordinal and continuous variables.

With this motivation, we propose a novel method to estimate associations between multilevel ordinal and continuous data using a latent Gaussian copula model approach. We assume that the multilevel ordinal variable is obtained by discretizing a latent variable, and estimate the correlation/covariance matrix underlying the Gaussian copula model via a rank-based approach. These results extend those for the latent Gaussian copula model for binary and continuous data proposed by \citep{Fanetal2017}.

In the next section we review the concept of Gaussian copula model and define a new latent Gaussian copula model for ordinal-continuous mixed data and review the motivations for Kendall's rank correlation coefficient. In Section 3, we propose the rank-based estimation for ternary-continuous mixed data and then generalize it to the estimation for ordinal-continuous mixed data. We derive explicit formulas for the bridge functions that connect the Kendall’s $\tau^a$ of observed data to the latent correlation matrix for different combinations of data types. This requires derivation of new bridge functions, and those derivations are somewhat involved and more complex than in continuous/binary case. We then use these formulas to construct a rank-based estimator of the latent correlation matrix for the mixed data. The significant advantage of bridge function technique is that it allows to estimate the latent correlation structure of Gaussian copula without estimating marginal transformation functions. We also establish theoretical concentration bound results for the new rank-based estimators. In Section 4 we consider the case of tied data.  Simulation results are presented in Section 5. In Section 6 we give two real data analysis that highlight the our proposed techniques applies them to the construction graphical models for mixed (binary, continuous, and ordinal) data.  The first is a well example in the algorithmic fairness literature about ProPublica's journalistic investigation on the apparent biases of machine learning based predictive analytics tool, COMPAS, in recidivism risk assessment \citep{angwin2016machine}.  The second example is another well know example first analyzed by \cite{ByarGreen1980} and subsequently by \cite{Hunt1999} consisting of 12 mixed type measurements for prostate cancer patients who were diagnosed as having either stage 3 or 4 prostate cancer.  We conclude with some discussion in Section 7. The proofs of the main results are given in the appendix.

\section{2 Background}\label{background}

\subsection{2.1 Variations of the Gaussian copula model}\label{gaussian-copula-model-for-continuous-data}

In recent years, the Gaussian copula model has received a lot of attention due to the ability to relax the normality assumptions of a fully Gaussian model. Formally the Gaussian copula model is defined as follows \citep{XueZou2012, Liu2009,Liu2012}:

\(\mathbf{Definition \text{ }1}\) (Gaussian copula model). \label{def1}A random vector \(\mathbf{X} = (X_1, \ldots, X_d)^T\) follows Gaussian copula model if there exists a set of monotonically increasing transformation functions \(f = (f_j)_{j = 1}^d\), such that \(f(\mathbf{X}) = (f_1(X_1), \ldots, f_d(X_d))^T \sim N_d(\mathbf{0},\Sigma)\) with \(\text{diag}(\Sigma) = 1\).

A random vector \(\mathbf{X}\) with these properties is said to follow a nonparanormal distribution denoted by \(\text{NPN} (0,\Sigma,f)\). The distribution is much more flexible that a Gaussian model. In particular, individual components of \(\mathbf{X}\) can have skewed or even multimodal distributions. 

Note that the Gaussian copula model only applies to continuous data. We now extend the latent Gaussian copula model to ordinal-continuous mixed data. Following the notation in the binary-continuous mixed case we consider a mixed-data random vector as
\(\mathbf{X} = (\mathbf{X_1}, \mathbf{X_2})\), where \(\mathbf{X_1}\) is \(d_1\)-dimensional vector of \(p\)-level discrete variables (with each component of \(X_1\) taking values in \(\{0, 1,\ldots, p-1\}\)) and \(\mathbf{X_2}\) is a \(d_2\)-dimensional vector of continuous variables.

\(\mathbf{Definition \text{ }2}\) (Latent Gaussian copula models for ordinal-continuous data). The random vector \(\mathbf{X}\) follows the extended latent Gaussian copula model if there exists a \(d_1\)-dimensional random vector of latent variables \(\mathbf{Z_1} = (Z_1, \ldots, Z_{d_1})\) such that \(X_j = l\) if \(Z_j \in (C_j^{l}, C_j^{l+1})\) for \(l = 0,1, \ldots, p-1\) and \(j = 1, \ldots, d_1\), where the cutoff vector is given by \(\mathbf{C} = (\mathbf{C}_1, \ldots, \mathbf{C}_{d_1})\) and \(\mathbf{C}_j = (C_j^0 = -\infty, C_j^1, \ldots, C_j^{p-1}, C_j^p = \infty)\) is an increasing sequence of \((p-1)\) constants, and \(\mathbf{Z} = (\mathbf{Z}_1, \mathbf{X}_2) \sim \text{NPN}(\mathbf{0},\mathbf{\Sigma},f)\).

The latent Gaussian copula model for binary-continuous data is just a special case of the above latent Gaussian copula model with \(p = 2\). Alternatively, the binary case is retrieved if \(C_j^2=\infty\) for \(j=1,\ldots,d_1\). In fact, by setting \(C_j^{k_j}=\infty\), where \(2\leq k_j\leq p\) for \(j=1,\ldots,d_1\), we can handle situations with ordinal variables with differing numbers of levels. \cite{Fanetal2017} proposed the following latent Gaussian copula model as an extension to binary and mixed binary-continuous data:

\(\mathbf{Definition \text{ }3 }\) (Latent Gaussian copula model for binary-continuous mixed data) Consider a mixed-data random vector \(\mathbf{X} = (\mathbf{X_1}, \mathbf{X_2})\) where \(\mathbf{X_1}\) is a \(d_1\)-dimensional vector of binary variables and \(\mathbf{X_2}\) is a \(d_2\)-dimensional vector of continuous variables. Then \(\mathbf{X}\) follows a latent Gaussian copula model if there exists a \(d_1\)-dimensional random vector of latent variables \(\mathbf{Z_1} = (Z_1, \ldots, Z_{d_1})\) such that \(X_j = I(Z_j > C_j)\) for \(j = 1, \ldots, d_1\) where \(\mathbf{C} = (C_1, \ldots, C_{d_1})\) is a \(d_1\)-dimensional vector of constants, with \(\mathbf{Z} = (\mathbf{Z}_1, \mathbf{X}_2) \sim \text{NPN}(\mathbf{0},\mathbf{\Sigma},f)\).



Our interest is in estimating the correlation matrix \(\mathbf{\Sigma}\) or the  precision matrix \(\mathbf{\Omega}= \mathbf{\Sigma^{-1}}\) with for latent Gaussian copula models for ordinal-continuous data. Furthermore, under the Gaussian copula model, the sparsity pattern of the precision matrix \(\mathbf{\Omega}\) reveals the conditional dependencies between \(X_j's\) for \(j = 1,2,...,d\). Hence the graph structure could also be recovered by estimating \(\mathbf{\Sigma}^{-1}\) as in the prostate cancer diagnostic example in Section 6.2.

\subsection{2.2 Kendall's rank correlation coefficients}
Kendall's $\tau^a$ (Kendall's rank correlation coefficient) is a nonparametric measure of nonlinear dependence between two random variables. It is similar to Spearman’s $\rho$ and Pearson’s $r$, in that is measures the relationship between two variables. Even though Kendall's $\tau^a$ is a similar to Spearman’s $\rho$ in that it is a nonparametric measure of relationship it differs in the interpretation of the correlation value. Spearman’s $\rho$ and Pearson’s $r$ magnitude are similar, however, Kendall’s $\tau^a$ is the difference between the probability that the observed data are in the same order versus the probability that the observed data are not in the same order.

Suppose the data consists of \(n\) independent \(d\)-dimensional random vectors, \(\mathbf{X}_1,\ldots, \mathbf{X} _n\), from a latent Gaussian copula model. The rank-based estimation framework for \(\mathbf{\Sigma}\), depending on the data type.
Specifically, estimation is based on the ``bridge function'' that relates Kendall's \(\tau^a\) parameter, \(\tau_{jk}^a\), for each variable pair \((j,k)\), \(1<j<k<d\), with the correlation, \(\sigma_{jk}\), between them. Here, the parameter \(\tau_{jk}^a\) is given by
\begin{align}
\tau_{jk}^a = \mathbb{E} \bigg[\text{sign}\{ (X_{ij} - X_{i'j})(X_{ik} - X_{i'k})\}\bigg]\,,
\label{tau-a}
\end{align}
which can be estimated unbiasedly by the corresponding \(\tau^a\) statistic
\begin{align}
\hat{\tau}_{jk}^a = {{n}\choose{2}}^{-1}\sum\limits_{1 \leq i < i' \leq n} \text{sign}(X_{ij} - X_{i'j})(X_{ik} - X_{i'k})\,,
\label{tauhat-a}
\end{align}
or equivalently by
\begin{align} 
\hat{\tau}_{jk}^a=  \dfrac{C-D}{\binom{n}{2}}
\label{tauhat-a2}
\end{align}
where \(C\) and \(D\) are the number of concordant and discordant pairs
among \((X_{1j}, X_{1k}), \ldots, (X_{nj},X_{nk})\).

A variation of Kendall’s $\tau^a$ that accounts for the important case of ties is $\tau^b$. Binary and ordinal data are very likely to have a large number of ties in ranking and, as a result, Kendall's \(\tau^a\) is likely to under-estimate the sample correlation. Therefore we consider a modified version, known as Kendall's \(\tau^b\). 
\begin{align}
\hat{\tau}_{jk}^b = \hat{\tau}_{jk}^a\frac{{{n}\choose{2}}}{\sqrt{\big[\binom{n}{2} - t_{X_j}\big] \big[\binom{n}{2}- t_{X_k}\big]}} \label{taub}
\end{align}
where \(t_{X_j}=\sum\limits_{1\leq i < i'\leq n} I(X_{ij} = X_{i'j})\) is the number of pairs of tied values of the \(j\)th response, and similarly
\(t_{X_k}=\sum\limits_{1\leq i < i'\leq n} I(X_{ik} = X_{i'k})\).

Since Kendall's \(\tau^b\) is a ratio of random terms (and the denominator involves a square root), the population bridge function linking it to \(\sigma_{jk}\) is intractable. We therefor consider 1st-order and 2nd order Taylor series approximation instead of directly computing its expectation. However, we find there is almost no difference between the 1st- and 2nd-order Taylor series approximations, or between them and a Monte-Carlo approximation of the exact expectation..

\section{3 Methodology}\label{methodology}

Suppose the data consists of \(n\) independent \(d\)-dimensional random vectors, \(\mathbf{X}_1,\ldots, \mathbf{X} _n\), from a latent Gaussian copula model. In this section, we propose a rank-based estimation framework for \(\mathbf{\Sigma}\), depending on the data type. Specifically, estimation is based on the ``bridge function'' that relates Kendall's \(\tau^a\) parameter, \(\tau_{jk}^a\), for each variable pair \((j,k)\), \(1<j<k<d\), with the correlation, \(\sigma_{jk}\), between them. The main idea behind our alternative procedure is to exploit Kendall’s $\tau^a$ statistics to directly estimate the unknown correlation matrix, without explicitly  calculating the marginal transformation functions $f_j$. Recall that the Kendall $\tau^a$ statistics are invariant under monotonic transformations. For Gaussian random variables there is a one-to-one mapping between these two statistics. For Gaussian copula distributions Kendall’s $\tau^a$ is connected to the covariance matrix in Definition 1 by $\sigma_{jk}=\sin(\frac{\pi}{2} \tau^a_{ij})$.

\subsection{3.1 Estimate correlation between ternary and ternary
data}\label{estimate-correlation-between-ternary-and-ternary-data}

We begin by considering ternary (3-level) data, and then extend to the general \(p\)-level case in Section 3.3.  Now suppose \(\mathbf{X}_j\), \(j = 1,\ldots,d_1\) are discrete data with 3 categories, taking values \(\{0,1,2\}\). Then under latent Gaussian copula model, we have \(p = 3\), and the data are obtained by trichotomizing the latent variable \(Z_j\) at cutoffs \((C_j^1, C_j^2) ,\, C_j^1 < C_j^2\) such that \[X_{ij} = 
\begin{cases} 
0 & \text{if } f(Z_{ij}) \leq \Delta_j^1 \\
1 & \text{if } \Delta_j^1 < f(Z_{ij}) \leq \Delta_j^2 \\
2 & \text{if } f(Z_{ij}) > \Delta_j^2
\end{cases}\] where \(\Delta_j^l=f(C_j^l)\), for \(l=1,2\).

To estimate \(\mathbf{\Sigma}\), we divide this into 3 cases where: (i) for \(1 \leq j,k \leq d_1\), \(\sigma_{jk}\) is the correlation between ternary variables; (ii) for \(1 \leq j \leq d_1 <k \leq d\), \(\sigma_{jk}\) is the correlation between ternary and continuous variables; and (iii) for \(d_1 < j,k \leq d\), \(\sigma_{jk}\) is the correlation between continuous variables. In case (iii) it has been shown by \cite{Kendall1948} that \(r_{jk} = \sin\big(\frac{\pi}{2} \hat{\tau}_{jk}^a \big)\).
In the remainder of this section, we confine our attention to cases (i) and (ii) respectively.

We first consider Kendall's \(\tau^a\) for two tenary variables. There are only four cases that need to be considered in order to determine concordance and discordance: \[\begin{aligned}
&(X_{ij} \leq 1, X_{ik} \leq 1); \quad (X_{ij} \geq 1, X_{ik} \geq 1);\quad (X_{ij} \leq 1, X_{ik} \geq 1); \quad (X_{ij} \geq 1, X_{ik} \leq 1)\,.
\end{aligned}\] Combining the first two will give a concordant pair and combining the last two will give a discordant pair. So using equations (\(\ref{tau-a}\)) and (\(\ref{tauhat-a2}\)) we can directly calculate the ``bridge function'' between \(\tau_{jk}^a\) and \(\sigma_{jk}\) as
\begin{align}
\tau_{jk}^a &=\mathbb{P}(C) -\mathbb{P}(D)\nonumber \\
&= \mathbb{P}(X_{ij} \leq 1; X_{ik} \leq 1)\mathbb{P}(X_{i'j} \geq 1; X_{i'k} \geq 1) + \mathbb{P}(X_{ij} \geq 1; X_{ik} \geq 1)\mathbb{P}(X_{i'j} \leq 1; X_{i'k} \leq 1) \nonumber\\
&\quad\quad - \mathbb{P}(X_{ij} \leq 1; X_{ik} \geq 1)\mathbb{P}(X_{i'j} \geq 1; X_{i'k} \leq 1) -\mathbb{P}(X_{ij} \geq 1; X_{ik} \leq 1)\mathbb{P}(X_{i'j} \leq 1; X_{i'k} \geq 1)\nonumber\\
&\text{(all the tied pairs cases in the first line will cancel out from those in the second line)}\\
&=2\Phi_2(\Delta_j^2,\Delta_k^2,\sigma_{jk})\Phi_2(-\Delta_j^1,-\Delta_k^1,\sigma_{jk})-2\bigg[\Phi(\Delta_j^2)-\Phi_2(\Delta_j^2,\Delta_k^1,\sigma_{jk})  \bigg]\bigg[\Phi(\Delta_k^2)-\Phi_2(\Delta_j^1,\Delta_k^2,\sigma_{jk})  \bigg]\,,
\label{poptt}
\end{align}
where the last step follows from
\[\begin{aligned}
	\mathbb{P}(X_{ij} \leq 1, X_{ik} \leq1) &= \mathbb{P}(f_j(Z_{ij}) \leq \Delta_j^2, f_k(Z_{ik}) \leq \Delta_k^2) =\Phi_2(\Delta_j^2,\Delta_k^2,\sigma_{jk});\\
	\mathbb{P}(X_{ij} \leq 1, X_{ik} \geq 1) &=\mathbb{P}(X_{ij} \leq 1) - \mathbb{P}(X_{ij} \leq 1, X_{ik} \leq 1) = \Phi(\Delta_j^2) -\Phi_2(\Delta_j^2,\Delta_k^2,\sigma_{jk}).
\end{aligned}\]
The notation \(\Phi_2(u,v,r)\) denotes the CDF of standard bivariate normal distribution with correlation \(r\), namely \(\Phi_2(u,v,r) = \int_{x_1< u}\int_{x_2 < v} \phi_2(x_1,x_2;r)dx_1dx_2\) where \(\phi_2(x_1,x_2;r)\) is the probability density function of the standard bivariate normal distribution with correlation \(r\).

It follows that the bridge function for the population Kendall's \(\tau^a\) for variable pair \((j,k)\), is given by \(\tau_{jk}^a = F(\sigma_{jk}; \Delta_j^1,\Delta_j^2,\Delta_k^1, \Delta_k^2)\) where

\begin{align}
F_a(\sigma_{jk}; \Delta_j^1,\Delta_j^2,\Delta_k^1, \Delta_k^2) &= 2\Phi_2(\Delta_j^2,\Delta_k^2,\sigma_{jk})\Phi_2(-\Delta_j^1,-\Delta_k^1,\sigma_{jk}) \nonumber \\
&\quad \quad -2\bigg[\Phi(\Delta_j^2)-\Phi_2(\Delta_j^2,\Delta_k^1,\sigma_{jk})  \bigg]\bigg[\Phi(\Delta_k^2)-\Phi_2(\Delta_j^1,\Delta_k^2,\sigma_{jk})  \bigg]\,.  \label{tt}
\end{align}

It will be shown in Lemma 3.1 that, for fixed \(\Delta_j^1, \Delta_j^2,\Delta_k^1, \Delta_k^2\), \(F_a(\sigma_{jk};\Delta_j^1, \Delta_j^2,\Delta_k^1, \Delta_k^2)\) is an invertible function of \(\sigma_{jk}\).

Simple moment estimators can be derived for the cutoffs using the relations \[\mathbb{E}(\mathbbm{1}\{X_{ij} = 0 \}) = \Phi(\Delta_j^1) \quad \text{ and } \quad \mathbb{E}(\mathbbm{1}\{X_{ij} = 2 \}) = 1-\Phi(\Delta_j^2)\,.\]
Specifically, these motivate the estimators \[\hat{\Delta}_j^1=\Phi^{-1} \bigg(\frac{\sum_i\mathbbm{1} \{X_{ij}=0\}}{n} \bigg)\quad \text{ and } \quad \hat{\Delta}_j^2=\Phi^{-1} \bigg(1- \frac{\sum_i\mathbbm{1} \{X_{ij}=2\}}{n} \bigg).\]
Thus a rank-based estimator of \(\sigma_{jk}\) is given by
\begin{align}
\hat{R}_{jk} = F_a^{-1}(\hat{\tau}_{jk}^a; \hat{\Delta}_j^1, \hat{\Delta}_j^2,\hat{\Delta}_k^1, \hat{\Delta}_k^2).
\end{align}

As will be seen from the following lemma, the bridge function \(F_a(\sigma_{jk};\hat{\Delta}_j^1, \hat{\Delta}_j^2,\hat{\Delta}_k^1, \hat{\Delta}_k^2)\) is strictly increasing in \(\sigma_{jk}\), thus there exists a unique root for the equation \(F_a(\sigma_{jk};\hat{\Delta}_j^1, \hat{\Delta}_j^2,\hat{\Delta}_k^1, \hat{\Delta}_k^2) = \hat{\tau}_{jk}^a\) which can be efficiently solved by Newton's method.

\(\mathbf{Lemma\ 3.1}\) For any fixed \(\Delta_j^1, \Delta_j^2,\Delta_k^1, \Delta_k^2\), \(F_a(r;\Delta_j^1, \Delta_j^2,\Delta_k^1, \Delta_k^2)\) in equation \(\ref{tt}\) is a strictly increasing function on \(r \in (-1,1)\). Thus, the inverse function \(F_a^{-1}(\tau^a; \Delta_j^1, \Delta_j^2,\Delta_k^1, \Delta_k^2)\) exists.

The proof of Lemma 3.1 is given in Appendix A.1.

We note here that the bridge functions for the binary-ternary and binary-binary cases can be derived directly from (\ref{tt}), by setting \(\Delta_j^2=\infty\) and both \(\Delta_j^2=\infty\) and \(\Delta_k^2=\infty\) respectively. Using the identities \(\Phi_2(\infty,v,r)=\Phi(v)\), \(\Phi_2(u,\infty,r)=\Phi(u)\), and \(\Phi_2(-u,-v,r)=1-\Phi(u)-\Phi(v)+\Phi_2(u,v,r)\), we find
\begin{align}
F_a(\sigma_{jk}; \Delta_j^1,\infty,\Delta_k^1, \Delta_k^2) =
2\Phi_2(\Delta_j^1,\Delta_k^2,\sigma_{jk})\left( 1-\Phi(\Delta_k^1) \right)
-2\Phi(\Delta_k^2)\left(\Phi(\Delta_j^1)-\Phi_2(\Delta_j^1,\Delta_k^1,\sigma_{jk})\right)\,,\label{bt}
\end{align}
and
\begin{align}
F_a(\sigma_{jk}; \Delta_j^1,\infty,\Delta_k^1, \infty) =
2\left(\Phi_2(\Delta_j^1,\Delta_k^1,\sigma_{jk})-\Phi(\Delta_j^1)\Phi(\Delta_j^2)\right)\,,\label{bb}
\end{align}
the latter agreeing with equation (3) of \cite{Fanetal2017}.

\subsection{3.2 Estimate correlation between ternary and continuous data}\label{estimate-correlation-between-ternary-and-continuous-data}

We now consider the random vector pairs \((X_{ij}, X_{ik})\) where variable \(j\) is ternary and variable \(k\) continuous. The latent Gaussian copula model assumptions imply that the corresponding latent vector pairs, \((Z_{ij}, X_{ik})\), satisfy \[(U_{ij},V_{ik})\equiv (f_j(Z_{ij}), f_k(X_{ik})) \sim N\left(\begin{bmatrix}
0\\
0
\end{bmatrix},\begin{bmatrix} 1& {\Sigma}_{jk} \\ {\Sigma}_{jk} &1 \end{bmatrix}\right)\] independently, for \(i = 1, \ldots, n\), where \(\sigma_{jk}\) is the correlation between \(U_{ij}\) and \(V_{ik}\).

It follows that
\[(U_{ij}, U_{i'j}, \frac{V_{ik}-V_{i'k}}{\sqrt{2}})^T \sim N_3 \bigg( \begin{bmatrix}
0\\
0\\
0
\end{bmatrix}, \begin{bmatrix}
1 & 0 & \sigma_{jk}/\sqrt{2} \\
0 & 1 & -\sigma_{jk}/\sqrt{2} \\
\sigma_{jk}/\sqrt{2} &  -\sigma_{jk}/\sqrt{2} &1 
\end{bmatrix}
\bigg).\]

Let \(\Phi_3\) denote the CDF for \((U_{ij}, U_{i'j}, \frac{V_{ik}-V_{i'k}}{\sqrt{2}})\),\\
\begin{align}
\Phi_3(a,b,c) =\mathbb{P} (U_{ij}<a, U_{i'j}<b, \frac{V_{ik}-V_{i'k}}{\sqrt{2}} < c). \label{phi3}
\end{align}

Now we are ready to build the bridge function of the population Kendall's \(\tau^a\) for ternary and continuous variables as follows.

\(\mathbf{{Lemma\ 3.2}}\) When \(X_{ij}\) is ternary and \(X_{ik}\) is continuous, \(\tau_{jk}^a = E(\hat{\tau}_{jk}^a)\) is given by \(\tau_{jk}^a = F_a(\sigma_{jk}; \Delta_j^1, \Delta_j^2)\) where
\begin{align}
 F_a(r; \Delta_j^1,\Delta_j^2) = 4\Phi_2 (\Delta_j^2,0,r/\sqrt{2}) -2 \Phi(\Delta_j^2) + 2[\Phi_3(\Delta_j^1, \Delta_j^2,0) - \Phi_3(\Delta_j^2,\Delta_j^1,0)]\,. \label{tc}
\end{align}

The next lemma shows that, for fixed \(\Delta_j^1, \Delta_j^2\), \(F(r;\Delta_j^1, \Delta_j^2)\) is an invertible function of \(r\), which implies that the equation has unique solution \(\hat{r} = F_a^{-1}(\hat{\tau}^a_{jk};\hat{\Delta}_j^1, \hat{\Delta}_j^2)\) where the unknown cut-offs \(\Delta_j^1, \Delta_j^2\) can be estimated with no bias by considering their expectations: \(\hat{\Delta}_j^1=\Phi^{-1} \bigg(\frac{\sum_i\mathbbm{1} \{X_{ij}=0\}}{n} \bigg)\) and \(\quad \hat{\Delta}_j^2=\Phi^{-1} \bigg(1- \frac{\sum_i\mathbbm{1} \{X_{ij}=2\}}{n} \bigg).\)

\(\mathbf{{Lemma\ 3.3}}\) For any fixed \(\Delta_j^1, \Delta_j^2\), \(F(r;\Delta_j^1, \Delta_j^2)\) in equation \(\ref{tc}\) is a strictly increasing function on \(r \in (-1,1)\) . Thus, the inverse function \(F_a^{-1}(\tau^a; \Delta_j^1, \Delta_j^2)\) exists.

The proofs of Lemma 3.2 and the following Lemma 3.3 can be found in Appendices A.2 and A.3.

Combining all three lemmas, we have constructed the rank-based estimate of \(\mathbf{\Sigma}\) as follows:

\[r_{jk} = \begin{cases} 
F_a^{-1}(\hat{\tau}_{jk}^a; \hat{\Delta}_j^1, \hat{\Delta}_j^2,\hat{\Delta}_k^1, \hat{\Delta}_k^2) & \text{for} 1 \leq j,k \leq d_1 \\
F_a^{-1}(\hat{\tau}_{jk}^a; \hat{\Delta}_j^1, \hat{\Delta}_j^2) & \text{for } 1 \leq j \leq d_1 <k \leq d \\
 \sin\big(\frac{\pi}{2} \hat{\tau}_{jk}^a \big) & \text{for } d_1 < j,k \leq d
\end{cases}\]

\subsection{\texorpdfstring{3.3 Generalized rank-based estimate for \(p\)-level discrete-continuous mixed data}{3.3 Generalized rank-based estimate for p-level discrete-continuous mixed data}}\label{generalized-rank-based-estimate-for-p-level-discrete-continuous-mixed-data}

We now generalize the rank-based estimate to \(p\)-level discrete-continuous mixed data. Suppose that \(\mathbf{X}_{j}\) is a \(p\)-level ordinal variable and \(\mathbf{X}_{k}\) continuous, then the bridge function for \(p\)-level discrete-continuous mixed data is established in the following lemma:

\(\mathbf{{Lemma\ 3.4}}\) When \(X_{ij}\) is \(p\)-level discrete taking value in \(\{0,1,\ldots,p-1\}\), and \(X_{ik}\) is continuous, the population version of Kendall's \(\tau^a\) is given by \(\tau_{jk}^a = F_a(\sigma_{jk};\mathbf{\Delta_j})\), where
\begin{align}
F_a(\sigma_{jk}; \mathbf{\Delta_j}) &= \sum\limits_{l=1}^{p-1} 4\Phi_3(\Delta_j^l, \Delta_j^{l+1},0)- 2\Phi(\Delta_j^l)\Phi(\Delta_j^{l+1}). \label{p-level}
\end{align}

Moreover, if we consider the entire \(\mathbb{Z}^+\) space, we can extend the estimates \(\mathbf{\hat{\Delta}_j}\) as 
\[ \hat{\Delta}_j^l = \Phi^{-1} (\frac{\sum_{i=1}^n I(X_{ij} \leq l - 1)}{n})\quad \quad \text{for } l \in \mathbb{Z}^+  \] 
so that for \(p\)-level mixed data ranging from \(0, \ldots,p-1\) then for \(l > p\) we have \(\hat{\Delta}_j^l= \infty\). If we define \(\Delta_j^l=\infty\) for \(l>p\), then \(4\Phi_3(\Delta_j^l, \Delta_j^{l+1},0) - 2\Phi(\Delta_j^l)\Phi(\Delta_j^{l+1}) = 4\times \frac{1}{2} - 2\times 1 \times 1 = 0.\)
Therefore we can write the \(\infty\)-form bridge function as:
\[F_a(r; \mathbf{\Delta_j}) = \sum\limits_{l = 1}^\infty 4\Phi_3(\Delta_j^l, \Delta_j^{l+1},0) - 2\Phi(\Delta_j^l)\Phi(\Delta_j^{l+1}).\]

\subsection{\texorpdfstring{3.4 Theoretical results}{3.4 Theoretical Results}}

We now are ready to establish a theoretical result concerning the convergence rate of the correlation estimate. As mentioned in \cite{Fanetal2017}, these two assumptions impose little restrictions in practice.

{\emph Assumption 1:} (bounded correlations) There is a constant
\(\delta \geq 0\) such that \(|\sigma_{jk}| \leq 1-\delta\) for
\(1 \leq j < k \leq d\).

{\emph Assumption 2:} (bounded cut-offs) There is a constant \(M\) such that
\(|\Delta_j^1| \leq M\) and \(|\Delta_j^2| \leq M\) for any
\(j = 1, \ldots, d\).

In the case of the estimate of correlation between $p$-level ordinal and continuous data we have the following concentration result.

\(\mathbf{Theorem\ 3.1}\) Under Assumptions 1 and 2, at fixed \(p\), for any \(t>0\) we have the following property \[P \big(\big|r_{jk} - \sigma_{jk}|> t \big) \leq 4\exp\bigg(-\frac{2M^2 n}{L_1^2} \bigg) + 2\exp\big(-\frac{nt^2}{2L_4^2} \big) + 2\exp( - \frac{nt^2\pi}{48^2 L_1^2 L_4^2 }) + 2\exp( - \frac{nt^2\pi}{24^2 L_1^2 L_4^2 })\]
implying that with probability greater than \(1- d^{-1}\)
\[\sup\limits_{1 \leq j < k \leq d} ||r_{jk} - \sigma_{jk}||  < C\sqrt{\frac{\log dp}{n}}\]
where \(L_1\) and \(L_4\) are positive constants defined in Appendix A9.
Essentially, Theorem 3.1 implies that for some constant \(\omega\)
independent of \(n\) and \(d\),
\(\sup\limits_{1\leq j < k \leq d} |r_{jk} - \sigma_{jk}| \leq \omega \sqrt{(\log d)/n}\)
with probability \(1-d^{-1}\).

We have a similar concentration rate for the correlation estimator of ternary-continuous mixed data.

\(\mathbf{Corollary\ 3.1}\) Under assumptions 1 and 2, for any \(t>0\) we
have
\[P \big(\big|r_{jk} - \sigma_{jk}\big| > t \big) \leq 4\exp\bigg(-\frac{2M^2 n}{L_1^2} \bigg) + 2\exp\big(-\frac{nt^2}{2L_4^2} \big) + 2\exp( - \frac{nt^2\pi}{48^2 L_1^2 L_4^2 }) + 2\exp( - \frac{nt^2\pi}{24^2 L_1^2 L_4^2 }).\]

\section{\texorpdfstring{4 Kendall's \(\tau^b\) for tied data}{4 Kendall's \textbackslash{}tau\^b}}\label{kendalls-taub}

Here we propose another correlation estimate for binary data as a
variant to the one proposed by \cite{Fanetal2017}, with the bridge
function given by (\ref{bb}).

\subsection{\texorpdfstring{4.1 Kendall's \(\tau^b \) estimate for
binary and binary
variables}{4.1 Kendall's \textbackslash{}tau\^{}b estimate for binary and binary variables}}\label{kendalls-taub-estimate-for-binary-and-binary-variables}

\(\mathbf{{Lemma\ 4.1}}\) When \(X_{ij}\) and \(X_{ik}\) are both binary
discrete random variables, the 1st-order Taylor series approximation of
the population version of Kendall's \(\tau^b\), given by
\(\tau_{jk}^b = E(\hat{\tau}_{jk}^b)\), is
\[F_b (\sigma_{jk};\Delta_j,\Delta_k) = \dfrac{\Phi_k(\Delta_j,\Delta_k,\sigma_{jk}) -\Phi(\Delta_j) \Phi(\Delta_k) }{\sqrt{(\Phi(\Delta_j)- \Phi(\Delta_j)^2)(\Phi(\Delta_k)- \Phi(\Delta_k)^2)}}.\]
We can easily see that \(F_b (\sigma_{jk};\Delta_j,\Delta_k)\) is
strictly increasing in \(\sigma_{jk}\) since the denominator is
independent of \(\sigma_{jk}\) and the numerator is the bridge function
for Kendall's \(\tau^a\). Therefore the equation
\(r_{jk} = F_b^{-1} (\hat{\tau}_{jk}^b;\hat{\Delta}_j,\hat{\Delta}_k)\)
has a unique solution.

\subsubsection{\texorpdfstring{4.2 Kendall's \(\tau^b\) estimate for
binary and continuous
variables}{4.2 Kendall's \textbackslash{}tau\^{}b estimate for binary and continuous variables}}\label{kendalls-taub-estimate-for-binary-and-continuous-variables}

\(\mathbf{{Lemma\ 4.2}}\) When \(X_{ij}\) is binary and \(X_{ik}\) is
continuous, the 1st-order Taylor series approximation of the population
version of Kendall's \(\tau^b\), given by
\(\tau_{jk}^b = E(\hat{\tau}_{jk}^b)\), is
\[F_b (\sigma_{jk};\Delta_j) = \dfrac{4\Phi_2(\Delta_j, 0, \sigma_{jk}/\sqrt{2})-2\Phi(\Delta_j)}{\sqrt{2(\Phi(\Delta_j))- 2(\Phi(\Delta_j))^2}}.\]
This bridge function is also strictly increasing in
\(\sigma_{jk} \in (-1,1)\) because the denominator does not involve
\(\sigma_{jk}\) and the numerator has been shown to be monotonically
increasing by \cite{Fanetal2017}.

We also derived the 2nd-order Taylor approximation of the bridge
function in this case.

\(\mathbf{{Lemma\ 4.3}}\) Let \(T_j=\sqrt{{{n}\choose{2}}-t_{X_j}}\).
Then, the 2nd-order Taylor approximation of \(E(\hat{\tau}_{jk}^b)\) is
given by

\[\begin{aligned}
\mathbb{E}(\hat{\tau}_{jk}^b)
\approx \frac{\sqrt{\binom{n}{2}}\mathbb{E}(\hat{\tau}_{jk}^a)}{\mathbb{E}(T_j)}+\left[\mathbb{E}(T_j)\right]^{-2}\bigg[\binom{n}{2}\text{var}(T_j)\frac{\sqrt{\binom{n}{2}}\mathbb{E}[\hat{\tau}_{jk}^a]}{\mathbb{E}(T_j)}
-\text{cov}\big(\sqrt{\binom{n}{2}}\hat{\tau}_{jk}^a,T_j\big)\bigg]
\end{aligned}\] where \[\begin{aligned}
\mathbb{E}[\hat{\tau}_{jk}^a] &= 4\Phi_2(\Delta_j, 0, r/\sqrt{2})-2\Phi(\Delta_j);\\
\mathbb{E}(T_j) &=\sum_{n_0=0}^n \bigg[\sqrt{\binom{n}{2} - \binom{n_0}{2} - \binom{n-n_0}{2}}\bigg] \binom{n}{n_0}\big(\Phi(\Delta_j)\big)^{n_0}\big(1-\Phi(\Delta_j)\big)^{n-n_0};\\
\quad   &\quad\\
\text{var}(T_j) &= \binom{n}{2}\big(2\Phi(\Delta_j) - 2[\Phi(\Delta_j)]^2 \big) - \mathbb{E}(T_j)^2;\\
\text{cov}\bigg(\sqrt{\binom{n}{2}}\tau_{jk}^a,T_j\bigg) &= \sum_{(C,D) \in S}\bigg\{ (C-D)\sqrt{(C+D)}\frac{\sqrt{\binom{n}{2}}}{C!D!\bigg(\binom{n}{2}-C-D \bigg)!}\cdot\\
&\quad\quad\quad\quad p_C^C p_D^D (1-p_C-p_D)^{\binom{n}{2}-C-D}\bigg\}- \sqrt{\binom{n}{2}}\mathbb{E}(\hat{\tau}_{jk}^a)\mathbb{E}(T_j)
\end{aligned}\] with the sample space of \((C,D)\) being
\(S = \{(C,D) : C\in \mathbb{Z}^+, D \in \mathbb{Z}^+, C+D \leq n \}\),
the probability of concordance and discordance respectively as
\(p_C = 2(\Phi_2(\Delta_j, 0, \sigma_{jk}/\sqrt{2}) - \Phi_3(\Delta_j, \Delta_j, 0))\)
and
\(p_D = 2(\Phi_2(\Delta_j, 0, -\sigma_{jk}/\sqrt{2}) - \Phi_3(\Delta_j, \Delta_j, 0))\).

\begin{figure}[htbp]
\centering
\includegraphics[height=3.12500in]{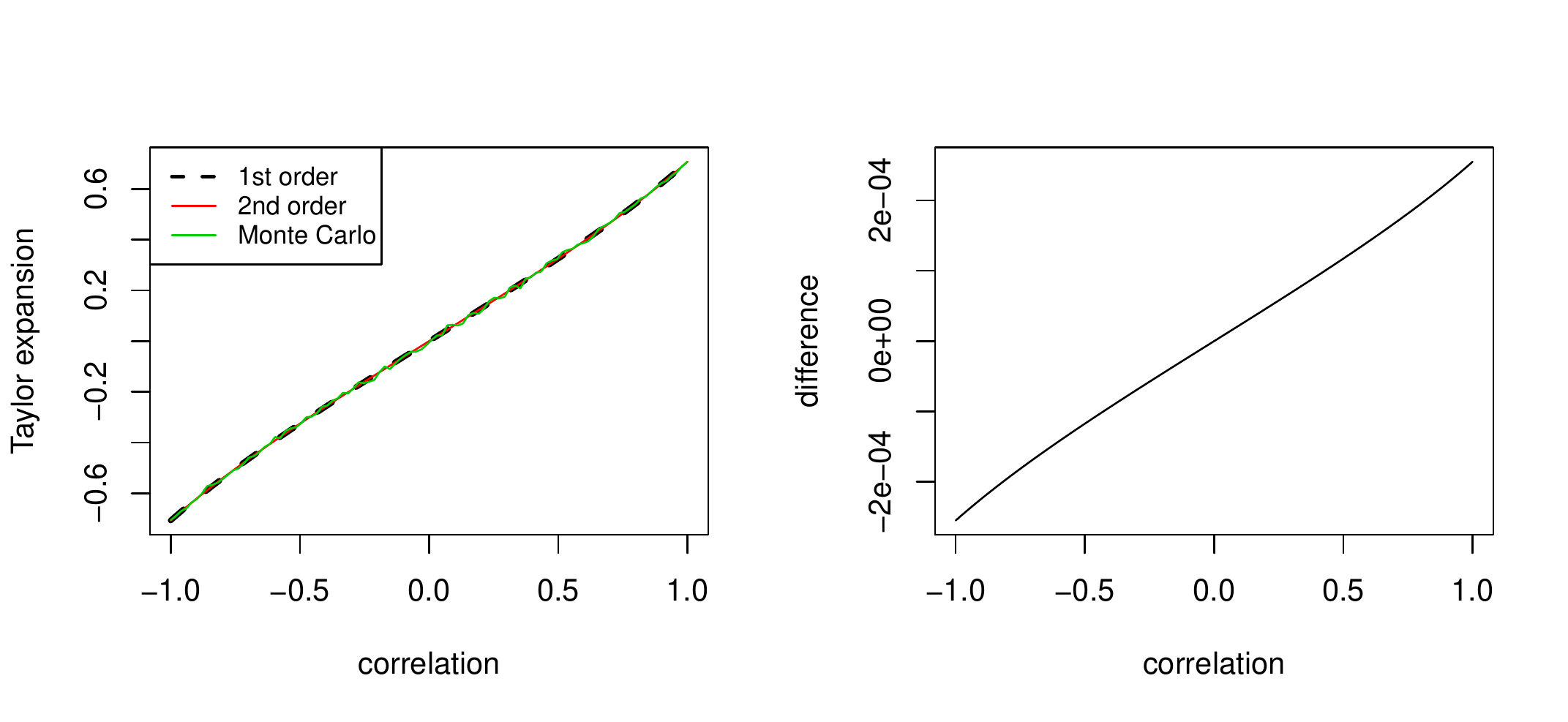}
\caption{Left: 1st-order and 2nd-order Taylor approximations visually
overlap with Monte-carlo simulated averages. Right: the difference
between 1st-order and 2nd-order Taylor approximations are negligible.}
\label{mcmc}
\end{figure}

The 1st-order and 2nd-order Taylor and Monte Carlo approximations to \(\tau^b\) are plotted in Figure \ref{mcmc} (right panel) for \(n=84\), and \(\Delta_j=0\). The difference between the two Taylor approximations is shown in the right panel.

\section{\texorpdfstring{5 Simulation results for generalized
\(p\)-level mixed
data}{4 Simulation results for generalized p-level mixed data}}\label{simulation-results-for-generalized-p-level-mixed-data}

In this section, we show some simulation results for \(p\)-level mixed
data where \(p = 2,3,\ldots, 16\). We conducted two scenarios here:

{\emph Scenario 1:} Starting with \(p = 2\), we dichotomize the data equally
by setting the cutoff \(\Delta_j = 0\). With \(p\) increasing, we
discretize the data by setting the cutoff
\(\Delta_j^l = \Phi^{-1}(1/p)\) so that we will have equal counts of
each level.

{\emph Scenario 2:} Starting with \(p = 16\), we discretize the continuous
Gaussian copula data equally so that each level has about the same
number of counts. As \(p\) decreases, we combine the highest level with
one level lower: e.g.~when \(p = 15\), we collapse ``16''s into ``15''s.
The motivation is that in Genetics research, when encountering ternary
data, people sometimes combine ``1''s and ``2''s to make the data
binary. As we can see in the following plot, this will lead to an
increased estimation error (see leftmost plots in Figure 2).

For each scenario, we first simulate bivariate Gaussian copula data of
size \(n=100\), \(d = 2\) and \(f(x) = x\), with the
correlation/covariance \(r =\{0, 0.01,\ldots, 1\}\), and we estimate the
correlations using the continuous data. Then we discretize the first
dimension of the data into \(p\) level in the way described by each
Scenario, and estimate the correlation following the bridge function in
equation (\(\ref{p-level}\)). For each \(r\), the same process is
repeated by 80 times and we take their mean of the squares as the error
measure. We further smooth the curve by averaging the errors over
\(r \in [0,0.1)\), \(r \in [0.1,0.2)\), etc.

We can see from the following plot that as \(p\) increases, the
estimation error approaches to the one in raw continuous data. However,
notice that how much estimation error will be introduced by combining
levels as we can see in Figure 2.

\begin{figure}[htbp]
\centering
\includegraphics{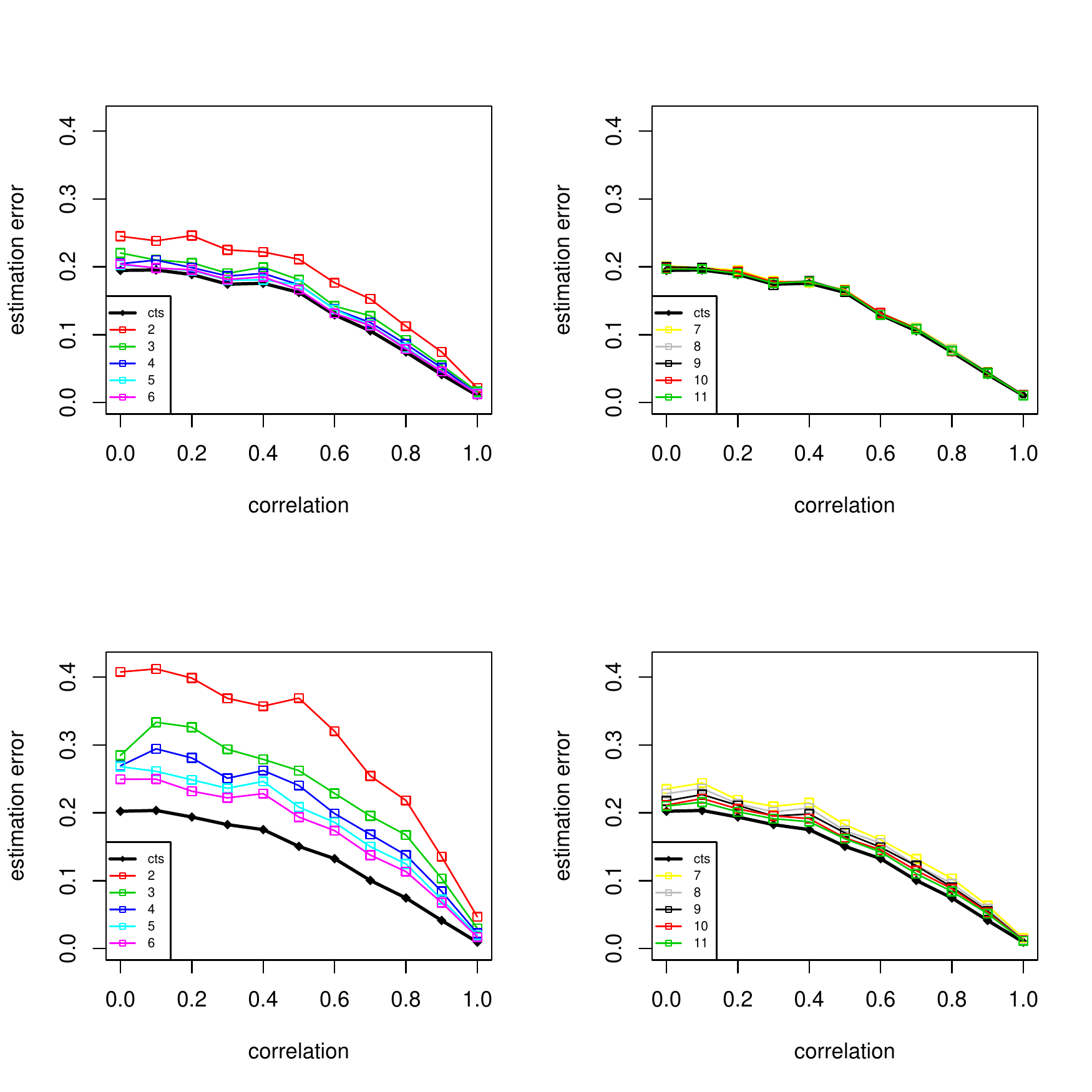}
\caption{Top: Simulation results for Scenario 1. For every \(p\), each
level of data has about the same size. As \(p\) increase, the estimation
error gets close to the one without discretizing the data.
Bottom:Simulation results for Scenario 2. For every \(p\), each level of
data has about the same size. As \(p\) increase, the estimation error
gets close to the one without discretizing the data.}
\end{figure}

\section{6 Real Data analysis}\label{real-data-analysis}
In this section, we present two studies of real data analysis. We start with applying our correlation estimation method to two sets of real data that have been studied intensively in the past, and then pass the correlation estimator to graph estimation procedures in next step. In the graph estimation procedure, we adopt the modified graphical lasso estimation method as in \cite{Fanetal2017}, which essentially consists of two steps: first we project the correlation estimator $\hat{\mathbf{R}}$ into the cone of positive semidefinite matrices to facilitate the optimization algorithms in \cite{Friedman08}, denoted as $\hat{\mathbf{R}}_p$; second we pass $\hat{\mathbf{R}}_p$ to the graphical lasso estimation to replace the sample covariance matrix, to obtain the following precision matrix estimator:
$$\hat{\mathbf{\Omega}} = \argmin_{\mathbf{\Omega} \succeq 0} \{\text{tr}(\hat{\mathbf{R}}_p\mathbf{\Omega}) - \log |\mathbf{\Omega}| + \lambda \sum_{j \neq k} |{\Omega}_{jk}| \}.$$
We set the path of tuning parameter to be the vector of length 10 starting from $\frac{\max{|\hat{\mathbf{R}}_p|}}{10}$ to $\max{|\hat{\mathbf{R}}_p|}$, as suggested by \cite{Friedman08}. Furthermore, we did not penalize the diagonal of inverse covariance matrix. We used high-dimensional BIC score (HBIC) as selection criterion, defined in \cite{Fanetal2017}.
The estimated graphs are then presented to reveal conditional independence relationships.

\subsection{6.1 COMPAS Data}\label{COMPAS}
ProPublica \citep{angwin2016machine} carried out a journalistic investigation on possible biases of machine learning based predictive analytic tools used in criminal justice.  The ProPublica article examined whether black-box risk assessment tools disproportionately recommend nonrelease of African-American defendants.  COMPAS (\underline{C}orrectional \underline{O}ffender \underline{M}anagement \underline{P}rofiling for \underline{A}lternative \underline{S}anctions) is a proprietary software tool developed by Northpointe, Inc. that gives a prediction score for a defendant's likelihood of failing to appear in court or reoffending. \cite{angwin2016machine} compiled criminal records from the criminal justice system in Broward County, Florida, combining detailed individual level criminal histories with predictions from the COMPAS risk assessment tool.  This data set has served as a key example in the algorithmic fairness literature (e.g.  \cite{adler2018auditing, berk2017fairness, chouldechova2017fair, johndrow2017algorithm, kleinberg2016inherent, tan2017detecting, zhou2018approximation}).  

The COMPAS score is computed by a black-box algorithm and produces a decile score (deciles of the predicted probability of rearrest) as well as a (ordinal) categorical score consisting of three levels of risk (low, medium, and high).  \cite{dieterich2016compas} suggest that a medium and high COMPAS scores garner more interest from supervision agencies than low scores.  In order to assess the accuracy of the recidivism predictions,  \cite{larson2016we} compared individual COMPAS score based predictions to a ground truth indicator of whether that particular individual had indeed been rearrested within two years of release.   \cite{larson2016we} developed a binary logistic regression model (low versus medium or high) that considered race, age, criminal history, future recidivism, and charge degree, they analyzed both the COMPAS scores for risk of overall and violent recidivism and used their model to assess the odds of getting a higher COMPAS score for certain subgroups.

\begin{figure}[htbp]
\centering
\includegraphics{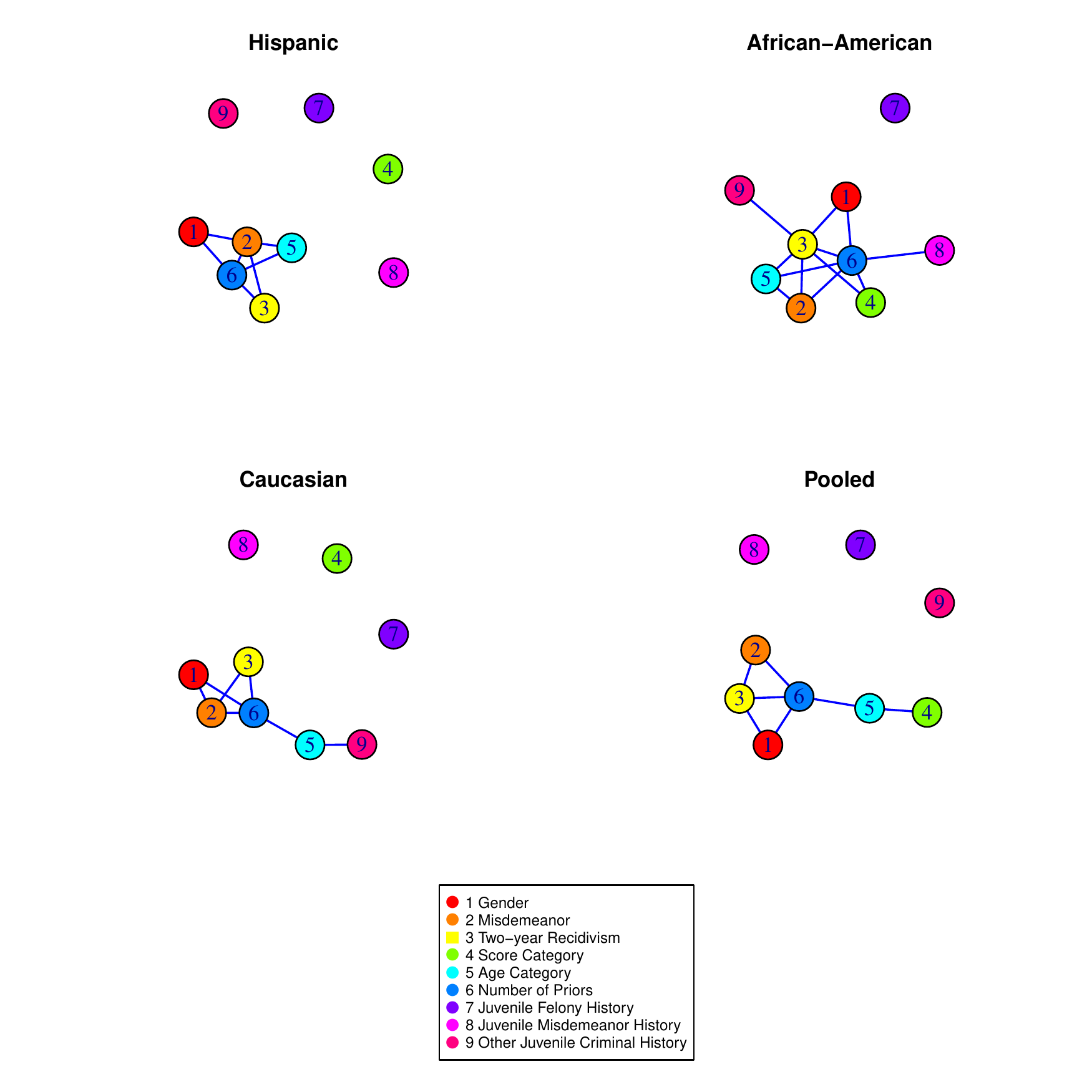}
\caption{Mixed data graphical model for the three level ordinal overall COMPAS score data set by African-American, Caucasian, Hispanic and pooled groups.}
\label{score}
\end{figure}

\begin{figure}[htbp]
\centering
\includegraphics{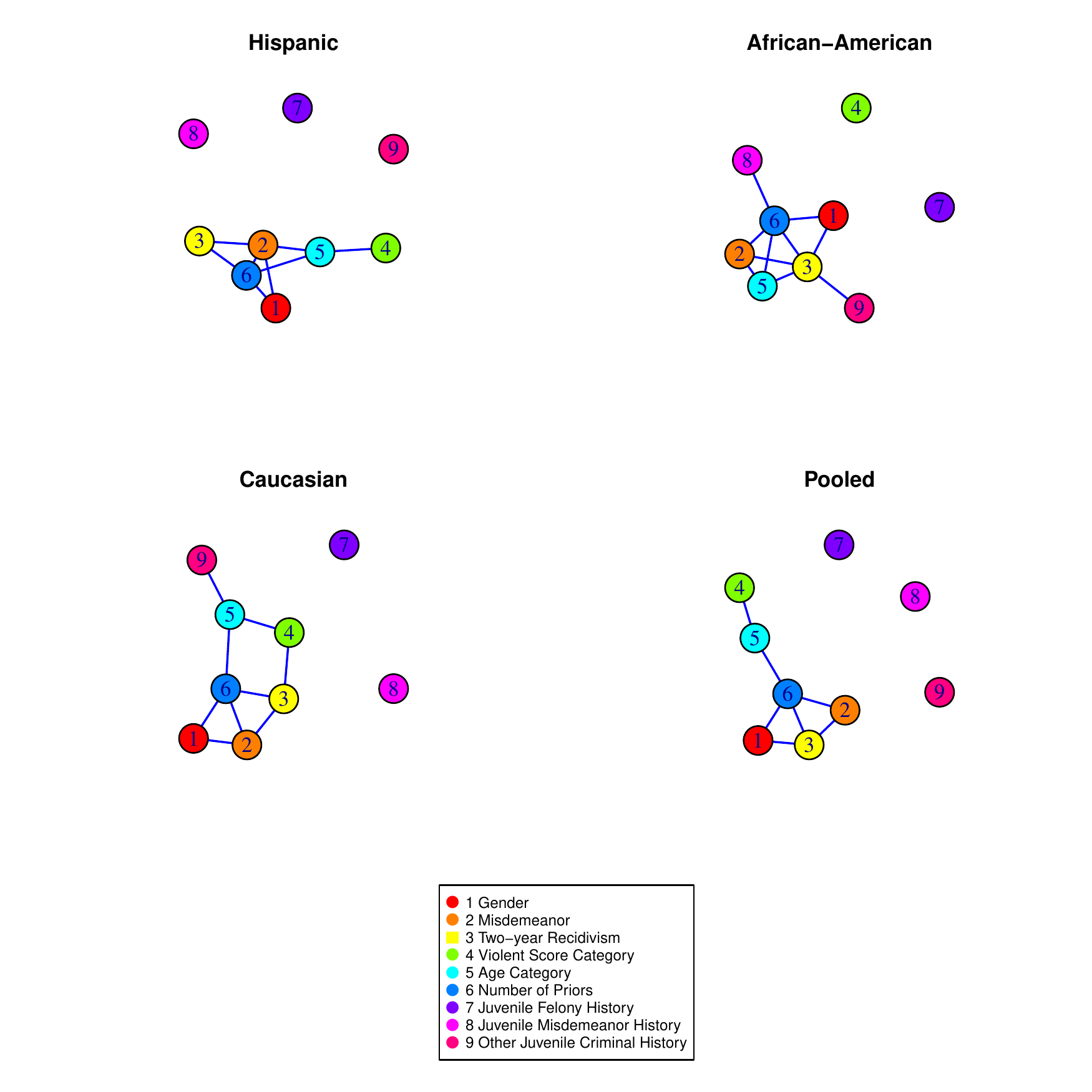}
\caption{Mixed data graphical model for the three level ordinal violent COMPAS score data set by African-American, Caucasian, Hispanic and pooled groups.}
\label{v_score}
\end{figure}

The ProPublica article \citep{angwin2016machine} mentions three African-Americans that had a medium risk COMPAS score and no subsequent offenses whereas non-African Americans had low risk score but had  subsequent serious offenses.  So it is of interest to examine the three level COMPAS score  (low, medium, and high) rather than the binary classification in \cite{larson2016we}.  We use our proposed graphical model approach to examine the conditional independence relationship between two-year recidivism (binary) and the three level (both overall and violent) COMPAS score  (low, medium, and high) with gender (binary), recorded misdemeanor (binary), age category (<25, 25-45, and >45), number of priors, and juvenile criminal history (felony, misdemeanor, and other -- all binary). To better understand the underlying relationships, we separate the data into three race groups: we estimated the underlying correlation matrices for African-American, Caucasian and Hispanic respectively, and also repeat the same procedure for the three races pooled together. Also, these analysis are done for overall COMPAS score categories and violent COMPAS score categories separately. Our estimated conditional independence graphs are in Figures \ref{score} and \ref{v_score}. A first interesting finding is that we notice the graphical structures vary across African-American, Caucasian and Hispanic groups.  In Figure \ref{score} for the overall COMPAS score, note that the overall COMPAS score has a direct effect on two-year recidivism for African-Americans but is conditionally independent for Caucasians and Hispanics, however has a quite indirect effect in the pooled model.  Conversely, in Figure \ref{v_score}, the violent COMPAS score is conditionally independent of two-year recidivism for African-Americans but has a direct effect for Caucasians and an indirect effect for Hispanics and the pooled groups.  It is also interesting how the various juvenile criminal history measures have different associations across the race groups and two COMPAS scores.  There is common structure seen across all three races too, misdemeanor and number of priors are consistently connected to two-year recidivism for all races. In contrast, this in not the case for misdemeanor and number of priors and the two COMPAS scores.  Also, the graphical models for pooled group are the same for both sets of variables involving score category and violent score category respectively.

\subsection{6.2 Prostate cancer data
analysis}\label{prostate-cancer-data-analysis}

\begin{figure}[htbp]
\centering
\includegraphics{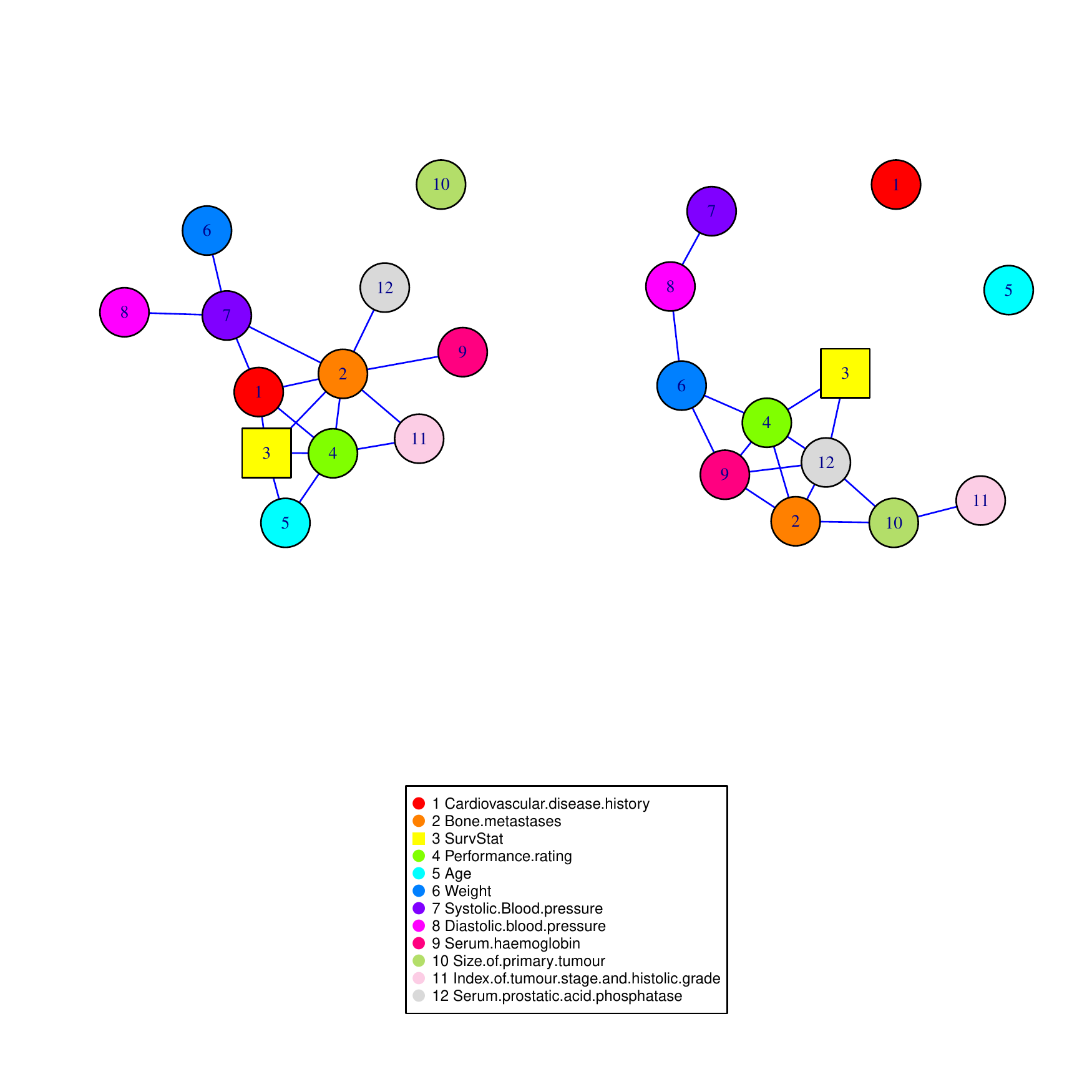}
\caption{Prostate cancer data analysis, Plot of the connected components
of the estimated graph for the prostate cancer data. Number 3 represents
the `Performance rating' variable. \textit{Left}: Stage 3 patients, the 'Survival Status' node is of degree 4, with neighbors including 'Cardiovascular disease history', ' Bone metastases', 'Performance Rating' and 'Age'; \textit{Right}: Stage 4 patients, 'Survival Status' node is of degree 2 instead, with its neighbors being 'Performance rating' and 'Serum prostatic acid phosphatase'.}
\label{prostate}
\end{figure}

This data set was first analyzed by \cite{ByarGreen1980} and subsequently by \cite{Hunt1999}. It consists of 12 mixed type measurements for 475 prostate cancer patients who were diagnosed as having either stage 3 or 4 prostate cancer. Among the 12 variables, 8 are continuous, 3 are ordinal and 1 is nominal (list of variables and corresponding abbreviations can be found in Table \ref{abbr}). More details of the data can be found at \citep{Andrews}. We are interested in how the `Survival Status' is correlated with the other 11 variables after removing the nominal variable `Electrocardiogram code' since it is not appropriate to infer latent variable for nominal variable. The 'Survival Status' is transformed into binary variable as either survived or died, regardless of causes of death. Also, we combined performance rating's level 2 and level 3 as one level since these patients are in bed more than 50\% of daytime. The correlation/covariance matrices are given in Table \ref{corr-table} and \ref{corr-table4} for Stage 3 and 4 patients respectively. Figure \ref{prostate} illustrates the recovered graph for the 12 variables for Stage 3 and Stage 4 patients respectively. It is interesting that the set of nodes connected to 'Survival Status' are different among Stage 3 and 4 patients. For Stage 3 patients, the 'Survival Status' node is of degree 3, with neighbors including 'Cardiovascular disease history' (HX), 'Bone metastases' (BM), and 'Performance Rating' (PF). Whilst for Stage 4 patients, 'Survival Status' node is of degree 2 instead, with its neighbors being 'Performance rating' and 'Serum prostatic acid phosphatase'. It is interesting that 'Performance rating' (PF) is adjacent to 'Survival Status' in both networks, which is reasonable since an active patient (Performance rating = 0 or 1) was probably able to move around hence survived. However, PF was not included in the best model found by \cite{Hunt1999}, which we speculate as a result of mistreating the categorical variable PF. Also, we notice that some variables are highly correlated with Surv but not a neighbor of Surv on the network graph, such as the 'Age' variable for Stage 3 patients, and Bone Metastases (BM) variable for Stage 4. It's easy to see the reason after a closer look at the correlation tables in Table \ref{corr-table} and \ref{corr-table4}: for Stage 3, the 'Age' variable has a higher correlation with 'PF' than with 'Surv', implying that the high correlation between 'Age' and 'Surv' might be a result of the high correlation between it and 'PF'. It is similar for Stage 4: 'BM' variable sees a higher correlation with 'PF' and 'AP', 'HG' has a higher correlation with 'PF' than with 'Surv', and 'SZ' finds itself highly correlated with 'AP' and 'PF', namely the high correlations between those variables and 'Surv' can be due to their high correlations with 'AP' and/or 'PF', thus they are indirectly connected to 'Surv' node in the network graph (Figure \ref{prostate}) but rather directly connected to the neighbors of 'Surv'. Another interesting structure can also be discovered from the network graph (Figure \ref{prostate}) that agrees with \cite{Hunt1999}: they found that the cluster consisting of variables 'BM', 'Wt', 'HG', 'SBP' and 'DBP' gave the second best likelihood; on the other hand, we found that those 5 variables are consistently clustered for both Stage 3 and 4 patients, which agrees with the finding by \cite{Hunt1999}. 
One might also notice that 'Size of primary tumor' node is isolated for only Stage 3 patients' network. This in fact agrees with the definition of Stage: stage 3 represents local extension of the disease whilst stage 4 represents distant metastasis as evidenced by elevated acid phosphatase and/or X-ray evidence \citep{Hunt1999}. In other words, for Stage 3 patients, 'SZ' (node 10) is not necessarily a good indicator of 'Index of tumor stage, histolic grade' (node 11) or 'Serum prostatic acid phosphatase' (node 12), but it might be a good one for stage 4 patients as we can see in the graph that node 10 is connected to node 11 and 12. 
Another interesting finding is that 'Size of primary tumor' and 'Serum prostatic acid phosphatase' are adjacent in the networks for Stage 4 patients, which agrees with the results in \cite{McParland} that Stage 4 patients on average saw larger tumors and higher levels of serum prostatic acid phosphatase. 

\begin{table}[ht]
\caption{List of variables and their abbreviations}
\centering
\begin{tabular}{|l|c|c|}
\hline
 Covariate & Abbreviation & Number of levels  \\ 
   &  & (if categorical) \\
 \hline 
  Cardiovascular.disease.history & HX & 2 \\ 
  Bone.metastases & BM & 2 \\ 
  SurvStat & Surv & 2 \\ 
  Performance.rating & PF & 3 \\ 
  Age & Age &   \\ 
  Weight & Wt &   \\ 
  Systolic.Blood.pressure & SBP &   \\ 
  Diastolic.blood.pressure & DBP &   \\ 
  Serum.haemoglobin & HG &   \\ 
  Size.of.primary.tumour & SZ &   \\ 
  Index.of.tumour.stage.and.histolic.grade & SG &   \\ 
  Serum.prostatic.acid.phosphatase & AP &   \\ 
  \hline
  \end{tabular}
\label{abbr}
  \end{table}


\begin{sidewaystable*}
	\small
	\setlength\tabcolsep{0pt}
	\caption{Correlation/covariance matrix for Stage 3 patients}
\begin{tabular*}{\textwidth}{ l @{\extracolsep{\fill}} *{26}{r} }
  \hline
 Variable & HX & BM & Surv & PF & Age & Wt & SBP & DBP & HG & SZ & SG & AP \\ 
  \midrule
HX & 1.00 & -1.00 & 0.48 & 0.39 & 0.27 & -0.01 & 0.24 & 0.09 & -0.09 & -0.07 & -0.17 & -0.17 \\ 
  BM & -1.00 & 1.00 & 1.00 & -1.00 & -0.09 & -0.14 & -0.67 & -0.03 & -0.76 & 0.06 & 0.55 & 0.91 \\ 
  Surv & 0.48 & 1.00 & 1.00 & 0.26 & 0.22 & -0.15 & 0.07 & 0.05 & -0.06 & 0.18 & 0.12 & -0.05 \\ 
  PF & 0.39 & -1.00 & 0.26 & 1.00 & 0.34 & -0.05 & 0.14 & 0.05 & -0.04 & -0.05 & 0.26 & 0.01 \\ 
  Age & 0.27 & -0.09 & 0.22 & 0.34 & 1.00 & 0.00 & 0.03 & -0.11 & -0.13 & -0.07 & -0.03 & -0.01 \\ 
  Wt & -0.01 & -0.14 & -0.15 & -0.05 & 0.00 & 1.00 & 0.23 & 0.18 & 0.17 & 0.06 & 0.06 & 0.12 \\ 
  SBP & 0.24 & -0.67 & 0.07 & 0.14 & 0.03 & 0.23 & 1.00 & 0.58 & 0.04 & 0.04 & -0.03 & -0.05 \\ 
  DBP & 0.09 & -0.03 & 0.05 & 0.05 & -0.11 & 0.18 & 0.58 & 1.00 & 0.14 & -0.05 & -0.05 & 0.01 \\ 
  HG & -0.09 & -0.76 & -0.06 & -0.04 & -0.13 & 0.17 & 0.04 & 0.14 & 1.00 & -0.06 & 0.07 & 0.17 \\ 
  SZ & -0.07 & 0.06 & 0.18 & -0.05 & -0.07 & 0.06 & 0.04 & -0.05 & -0.06 & 1.00 & 0.18 & 0.09 \\ 
  SG & -0.17 & 0.55 & 0.12 & 0.26 & -0.03 & 0.06 & -0.03 & -0.05 & 0.07 & 0.18 & 1.00 & 0.10 \\ 
  AP & -0.17 & 0.91 & -0.05 & 0.01 & -0.01 & 0.12 & -0.05 & 0.01 & 0.17 & 0.09 & 0.10 & 1.00 \\ 
   \hline
\end{tabular*}
\label{corr-table}\\

\medskip
\hfill
	\small
	\setlength\tabcolsep{0pt}
	\caption{Correlation/covariance matrix for Stage 4 patients}
\begin{tabular*}{\textwidth}{ l @{\extracolsep{\fill}} *{26}{r} }
  \hline
 Variable & HX & BM & Surv & PF & Age & Wt & SBP & DBP & HG & SZ & SG & AP \\ 
  \hline
HX & 1.00 & -0.07 & 0.15 & 0.16 & 0.16 & 0.12 & -0.02 & -0.10 & 0.06 & -0.09 & -0.05 & 0.01 \\ 
  BM & -0.07 & 1.00 & 0.33 & 0.50 & -0.07 & -0.29 & -0.07 & -0.12 & -0.42 & 0.28 & 0.11 & 0.33 \\ 
  Surv & 0.15 & 0.33 & 1.00 & 0.52 & 0.15 & -0.21 & 0.07 & -0.01 & -0.30 & 0.22 & 0.15 & 0.28 \\ 
  PF & 0.16 & 0.50 & 0.52 & 1.00 & 0.09 & -0.42 & 0.10 & -0.13 & -0.65 & 0.24 & 0.09 & 0.30 \\ 
  Age & 0.16 & -0.07 & 0.15 & 0.09 & 1.00 & -0.10 & 0.09 & -0.10 & -0.15 & 0.04 & 0.01 & 0.09 \\ 
  Wt & 0.12 & -0.29 & -0.21 & -0.42 & -0.10 & 1.00 & 0.15 & 0.25 & 0.36 & -0.04 & -0.11 & -0.15 \\ 
  SBP & -0.02 & -0.07 & 0.07 & 0.10 & 0.09 & 0.15 & 1.00 & 0.57 & 0.11 & 0.12 & -0.01 & -0.00 \\ 
  DBP & -0.10 & -0.12 & -0.01 & -0.13 & -0.10 & 0.25 & 0.57 & 1.00 & 0.17 & 0.04 & -0.03 & -0.09 \\ 
  HG & 0.06 & -0.42 & -0.30 & -0.65 & -0.15 & 0.36 & 0.11 & 0.17 & 1.00 & -0.15 & -0.09 & -0.20 \\ 
  SZ & -0.09 & 0.28 & 0.22 & 0.24 & 0.04 & -0.04 & 0.12 & 0.04 & -0.15 & 1.00 & 0.23 & 0.34 \\ 
  SG & -0.05 & 0.11 & 0.15 & 0.09 & 0.01 & -0.11 & -0.01 & -0.03 & -0.09 & 0.23 & 1.00 & 0.15 \\ 
  AP & 0.01 & 0.33 & 0.28 & 0.30 & 0.09 & -0.15 & -0.00 & -0.09 & -0.20 & 0.34 & 0.15 & 1.00 \\ 
   \hline
\end{tabular*}
\label{corr-table4}
\end{sidewaystable*}

\section{7 Conclusion and Discussion}\label{conclusion-and-discussion}

To sum up, we proposed a generalized rank-based method to estimate
correlations for any \(p\)-level discrete-continuous mixed data. The
method is under latent Gaussian copula model, assuming there is some
latent variable that discretize the continuous data into categorical.
There exists unique solution to the bridge function, which can be
obtained easily by Newton's method. The theoretical properties of the
estimates are well established. In our simulation studies, we see as
\(p\) increases, the estimation becomes as accurate as the one using raw
continuous data. This agrees with the intuition that as we obtain more
information, the estimation will do a better job.

Correlation estimates for ternary-ternary data and binary-ternary data
are also given, to help social science researches find associations
among different types of data.

Also, we proposed a modified estimate based on Kendall's \(\tau^b\)
compared to the one based on Kendall's \(\tau^a\) in \cite{Fanetal2017},
to account for occurrences of tied pairs. Since the Kendall's \(\tau^b\)
involves a square root term in its denominator, we did not compute its
population version directly but rather obtained its 1st-order and
2nd-order Taylor approximations, which showed no visible difference from
the Monte-Carlo simulated average.

Our method can further be applied to graph recovery by inverting the
correlation matrix estimate (into the so-called precision matrix).
Conditional independence can also be inferred from the precision matrix.
One practical advantage of our method is that it can estimate the
correlations regardless of dimensions. For high-dimensional data,
estimation can be done in parallel to reduce time expense.

\pagebreak

\section{Appendix}\label{appendix}

\subsection{A1. Proof of Lemma 3.1}\label{a1.-proof-of-lemma-2.1}

\begin{proof}
It is equivalent to show $\frac{\partial{ F(r;\Delta_j^1, \Delta_j^2,\Delta_k^1, \Delta_k^2)}}{\partial r} >0$. 

Let $\partial_{22}$ denote $\frac{\partial{ \Phi_2(\Delta_j^2,\Delta_k^2,r)}}{\partial r} >0$, and similar notations defined for $\partial_{11}$, $\partial_{12}$, $\partial_{21}$, then we have
$$\begin{aligned}
&\quad \quad \frac{\partial{ F(r;\Delta_j^1, \Delta_j^2,\Delta_k^1, \Delta_k^2)}}{\partial r}\\
&=\frac{\partial{\bigg[\Phi_2(\Delta_j^2,\Delta_k^2,r) - \Phi_2(\Delta_j^2,\Delta_k^2,r)\big(\Phi(\Delta_k^1)+\Phi(\Delta_j^1)  \big)+\Phi_2(\Delta_j^1,\Delta_k^1,r)\Phi_2(\Delta_j^2,\Delta_k^2,r)\bigg]}}{\partial r}\\
&= \partial_{22} - \Phi(\Delta_k^1) \partial_{22} -  \Phi(\Delta_j^1) \partial_{22} +\partial_{11}\Phi_2(\Delta_j^2,\Delta_k^2,r) + \partial_{22}\Phi_2(\Delta_j^1,\Delta_k^1,r)\\
&= [1-\Phi(\Delta_j^1) - \Phi(\Delta_k^1)+\Phi_2(\Delta_j^1,\Delta_k^1,r)\big]\partial_{22} +\partial_{11}\Phi_2(\Delta_j^2,\Delta_k^2,r) \\
&=\Phi_2(-\Delta_j^1,-\Delta_k^1,r)\partial_{22} +\partial_{11}\Phi_2(\Delta_j^2,\Delta_k^2,r)\\ 
&>0\\
&\quad \frac{\partial{[\big(\Phi_2(\Delta_j^2,\Delta_k^1,r)-\Phi(\Delta_j^2)  \big)\big(\Phi_2(\Delta_j^1,\Delta_k^2,r)-\Phi(\Delta_k^2)  \big)]}}{\partial r}\\
&=\big[\Phi(\Delta_k^2)- \Phi_2(\Delta_j^1,\Delta_k^2,r)\big]\partial_{21}-\big[\Phi_2(\Delta_j^2,\Delta_k^1,r)-\Phi(\Delta_j^2) \big]\partial_{12}\\
&=\mathbb{P}(U_{ij} > \Delta_j^1; V_{ik} <\Delta_k^2)\partial_{21}+\mathbb{P}(U_{ij} < \Delta_j^2; V_{ik} >\Delta_k^1)\partial_{12}\\
&>0.
\end{aligned}$$
\end{proof}

\subsection{A2. Proof of Lemma 3.2}\label{a2.-proof-of-lemma-2.2}

\begin{proof}

We know that 

$$\begin{aligned}
\text{sign} (X_{ij} - X_{i'j}) &= \mathbbm{1} \{X_{ij}=2\} - \mathbbm{1} \{X_{i'j}=2\}\\
&\quad \quad+\mathbbm{1} \{X_{ij}=1, X_{i'j} = 0 \} -\mathbbm{1} \{X_{ij}=0, X_{i'j} = 1 \}
\end{aligned}$$
thus it is true that 

$$\begin{aligned}
&\quad E[\text{sign} (X_{ij} - X_{i'j})(X_{ik} - X_{i'k})] \\
&= E\big[\mathbbm{1} \{X_{ij}=2\} \text{sign}(X_{ik} - X_{i'k})\big] - E\big[\mathbbm{1} \{X_{i'j}=2\} \text{sign}(X_{ik} - X_{i'k})\big]\\
&\quad \quad + E\big[\mathbbm{1} \{X_{ij}=1, X_{i'j} = 0 \} \text{sign}(X_{ik}- X_{i'k})\big] \\
& \quad \quad  - E\big[\mathbbm{1} \{X_{ij}=0, X_{i'j} = 1 \} \text{sign}(X_{ik} - X_{i'k})\big]
\end{aligned}.$$

We consider the four terms as two parts separately.
The first two terms can be further computed as
$$\begin{aligned}
&\quad E\big[\mathbbm{1} \{X_{ij}=2\} \text{sign}(X_{ik} - X_{i'k})\big] - E\big[\mathbbm{1} \{X_{i'j}=2\} \text{sign}(X_{ik} - X_{i'k})\big]\\
&= E\big[\mathbbm{1} \{U_{ij}> \Delta_j^2\} \text{sign}(X_{ik} - X_{i'k})\big] - E\big[\mathbbm{1} \{U_{i'j}> \Delta_j^2\} \text{sign}(X_{ik} - X_{i'k})\big]\\
&= 2 E\big[\mathbbm{1} \{U_{ij}> \Delta_j^2, V_{ik} - V_{i'k} >0 \}\big] - 2 E\big[\mathbbm{1} \{U_{i'j}> \Delta_j^2, V_{ik} - V_{i'k} >0 \}\big]\\
&= 2\Phi_2(\Delta_j^2,0,r/\sqrt{2}) - 2 \Phi_2(\Delta_j^2,0,-r/\sqrt{2})\\
&= 4\Phi_2 (\Delta_j^2,0,r/\sqrt{2}) -2 \Phi(\Delta_j^2).
\end{aligned}$$

The last two terms hold the following equivalence:
$$
\begin{aligned}
&\quad E\big[\mathbbm{1} \{X_{ij}=1, X_{i'j} = 0 \} \text{sign}(X_{ik}- X_{i'k})\big]- E\big[\mathbbm{1} \{X_{ij}=0, X_{i'j} = 1 \} \text{sign}(X_{ik} - X_{i'k})\big]\\
&=2E\big[\mathbbm{1} \{U_{ij} \in [\Delta_j^1,\Delta_j^2] , U_{i'j} <\Delta_j^1, V_{ik} - V_{i'k} >0  \} \big]\\
&\quad \quad - 2E\big[\mathbbm{1} \{U_{ij} <\Delta_j^1 , U_{i'j}\in [\Delta_j^1,\Delta_j^2] , V_{ik} - V_{i'k} >0\}\big]\\
&= 2[\Phi_3(\Delta_j^1, \Delta_j^2,0) - \Phi_3(\Delta_j^2,\Delta_j^1,0)].
\end{aligned}
$$

 Then we further have 
$$
\begin{aligned}
&\quad\quad2E\big[\mathbbm{1} \{U_{ij} \in [\Delta_j^1,\Delta_j^2] , U_{i'j} <\Delta_j^1, V_{ik} - V_{i'k} >0  \} \big] \\
&=2\bigg(\Phi_3(\Delta_j^2,\Delta_j^1, \infty) -\Phi_3(\Delta_j^1,\Delta_j^1,\infty) - \Phi_3(\Delta_j^2, \Delta_j^1, 0) +\Phi_3(\Delta_j^1,\Delta_j^1,0)  \bigg)\\
&= 2\bigg(\Phi(\Delta_j^1)\big(\Phi(\Delta_j^2) -\Phi(\Delta_j^1) \big)- \Phi_3(\Delta_j^2, \Delta_j^1, 0) +\Phi_3(\Delta_j^1,\Delta_j^1,0)\bigg),
\end{aligned}
$$

and likewise 
$$
\begin{aligned}
&\quad\quad2E\big[\mathbbm{1} \{U_{ij} <\Delta_j^1 , U_{i'j}\in [\Delta_j^1,\Delta_j^2] , V_{ik} - V_{i'k} >0\}\big] \\
&=2\bigg(\Phi_3(\Delta_j^1,\Delta_j^2, \infty) -\Phi_3(\Delta_j^1,\Delta_j^1,\infty) - \Phi_3(\Delta_j^1,\Delta_j^2, 0) +\Phi_3(\Delta_j^1,\Delta_j^1,0)  \bigg)\\
&= 2\bigg(\Phi(\Delta_j^1)\big(\Phi(\Delta_j^2) -\Phi(\Delta_j^1) \big)- \Phi_3(\Delta_j^1,\Delta_j^2, 0) +\Phi_3(\Delta_j^1,\Delta_j^1,0)\bigg).
\end{aligned}
$$

Hence the bridge function for ternary-continuous mixed data is found to be
$$\begin{aligned}
&\quad E[\text{sign} (X_{ij} - X_{i'j})(X_{ik} - X_{i'k})] \\
&= 4\Phi_2 (\Delta_j^2,0,r/\sqrt{2}) -2 \Phi(\Delta_j^2) + 2[\Phi_3(\Delta_j^1, \Delta_j^2,0) - \Phi_3(\Delta_j^2,\Delta_j^1,0)].
\end{aligned}$$

\end{proof}
\subsection{A3. Proof of Lemma 3.3}\label{a3.-proof-of-lemma-2.3}
\begin{proof}
We need to theoretically show the monotonicity of the bridge function
for ternary-continuous data which boils down to show the following it
monotonically increasing in \(r\):
\[ 4\Phi_3(\Delta_j^l, \Delta_j^{l+1},0) - 2\Phi(\Delta_j^l)\Phi(\Delta_j^{l+1})\]

hence it suffices to show that for all \(l\),
\(\dfrac{\partial \Phi_3(\Delta_j^{l-1},\Delta_j^l,0)}{\partial r} >0\)
Recall that the $\Phi_3$ is the cumulative distribution function for random variables $(U_{ij}, U_{i'j}, \frac{V_{ik}-V_{i'k}}{\sqrt{2}})^T$ as defined in Section \ref{estimate-correlation-between-ternary-and-continuous-data}, where $(U_{ij}, U_{i'j}, \frac{V_{ik}-V_{i'k}}{\sqrt{2}})^T \sim N_3 \bigg( \begin{bmatrix}
0\\
0\\
0
\end{bmatrix}, \begin{bmatrix}
1 & 0 & \sigma_{jk}/\sqrt{2} \\
0 & 1 & -\sigma_{jk}/\sqrt{2} \\
\sigma_{jk}/\sqrt{2} &  -\sigma_{jk}/\sqrt{2} &1 
\end{bmatrix}
\bigg).$ 

For easy notation, we denote $(U_{ij}, U_{i'j}, \frac{V_{ik}-V_{i'k}}{\sqrt{2}})^T$ as $\mathbf{x} = (x_1, x_2, x_3)^T$, and $$\Sigma = \begin{bmatrix}
1 & 0 & \sigma_{jk}/\sqrt{2} \\
0 & 1 & -\sigma_{jk}/\sqrt{2} \\
\sigma_{jk}/\sqrt{2} &  -\sigma_{jk}/\sqrt{2} &1 
\end{bmatrix}$$  for the rest of this proof.

Note that we can rewrite the normal density function $\phi_3(\mathbf{x},\Sigma)$ as the transform of its characteristic function \citep{Cramer46}:

\begin{align}
	\phi_3(\mathbf{x},\Sigma) &= (2\pi)^{-3}\iiint \exp(-i\mathbf{t}^T\mathbf{x} - \frac{1}{2} \mathbf{t}^T\mathbf{\Sigma}\mathbf{t})d\mathbf{t}\label{phi3pdf}
\end{align}

A result of this is 

\[\frac{\partial \phi_3(\mathbf{x})}{\partial r} = \big(\frac{\partial^2 \phi_3}{\partial x_1 \partial x_3} - \frac{\partial^2 \phi_3}{\partial x_2 \partial x_3}\big)\cdot (1/\sqrt{2})\]
which can be seen after interchanging the order of differentiation and integration in equation \ref{phi3pdf}. 

So we now have 
$$
\begin{aligned}
	&\quad\quad \frac{\partial \Phi_3(\Delta_j^l, \Delta_j^{l+1},0)}{\partial r/\sqrt{2}}\\
	&= \int_{-\infty}^{\Delta_j^l} \int_{-\infty}^{\Delta_j^{l-1}} \int_{-\infty}^0 \bigg[ \frac{\partial^2 \phi_3(x_1,x_2,x_3) }{\partial x_1 \partial x_3} - \frac{\partial^2 \phi_3(x_1,x_2,x_3)}{\partial x_2 \partial x_3}\bigg] dx_1 dx_2 dx_3 \\
	&= \int_{-\infty}^{\Delta_j^l} \int_{-\infty}^{\Delta_j^{l-1}} \bigg[\frac{\partial^2 \phi_3(x_1,x_2,0)}{\partial x_1 } - \frac{\partial^2 \phi_3(x_1,x_2,0)}{\partial x_2}\bigg] dx_1 dx_2\\
	&= \int_{-\infty}^{\Delta_j^l} \phi_3(\Delta_j^{l-1}, x_2, 0)dx_2 - \int_{-\infty}^{\Delta_j^{l-1}} \phi_3(x_1, \Delta_j^l, 0)dx_1.
\end{aligned}
$$
Recall that $(x_1, x_2, x_3) \stackrel{d}{=} (x_2, x_1, -x_3)$, we then have 
$$\begin{aligned}
&= \int_{-\infty}^{\Delta_j^l} \phi_3(\Delta_j^{l-1}, x, 0)dx - \int_{-\infty}^{\Delta_j^{l-1}} \phi_3(\Delta_j^{l}, x, 0)dx\\
&= \Phi(\Delta_j^l)\phi_2(\Delta_j^{l-1}, 0, r/\sqrt{2}) - \Phi(\Delta_j^{l-1})\phi_2(\Delta_j^l, 0, r/\sqrt{2})
\end{aligned}$$
where the last step arises from the fact that $X_2 | X_1 = \Delta_j^{l-1}, X_3 = 0 \sim N(0,1)$. 

Since $\Phi(\cdot) >0$, $\phi_2(\cdot, \cdot, r/\sqrt{2}) > 0$, so in order to show $\frac{\partial \Phi_3(\Delta_j^l, \Delta_j^{l+1},0)}{\partial r/\sqrt{2}} >0$, we only need to show $\frac{\Phi(\Delta_j^l)}{\phi_2(\Delta_j^l, 0, r/\sqrt{2})} > \frac{\Phi(\Delta_j^{l-1})}{\phi_2(\Delta_j^{l-1}, 0, r/\sqrt{2})},$
which is equivalent to show $\frac{ \Phi(y)}{\phi_2(y,0,r/\sqrt{2})}$ is increasing in $y$. Now we also notice that $\phi_2(y,0,r/\sqrt{2}) = \phi\big( \frac{y}{\sqrt{1 -\frac{ r^2}{2}}} \big)\phi(0)$ due to the conditional distribution property of bivariate normal variables. 
Therefore it is equivalent to show $\frac{\Phi(y)}{\phi\big( \frac{y}{\sqrt{1 - \frac{ r^2}{2}}} \big)}$ increasing in $x$. However, we know that $\frac{\Phi(y)}{\phi\big( \frac{y}{\sqrt{1 - \frac{ r^2}{2}}} \big)} = \frac{\Phi(-y)}{\phi\big(- \frac{y}{\sqrt{1 - \frac{ r^2}{2}}} \big)}$, 
Let $\lambda(y) = \frac{\Phi(y)}{\phi\big( \frac{y}{c} \big)}$ and $c =\sqrt{1 -\frac{ r^2}{2}}$ , then since $\phi'(x) = \phi(x)\cdot(-x)$ we have
$$\begin{aligned}
	\lambda'(y) &= \frac{\phi(y)\phi(\frac{y}{c}) - \Phi(y)\phi(\frac{y}{c})(-\frac{y}{c^2})}{\phi^2(y/c)} \\
	&= \frac{\phi(y)}{\phi(y/c)} - \frac{\Phi(y)}{\phi(y/c)}(-\frac{y}{c^2})\\
	&= \lambda(y)\bigg(\frac{\phi(y)}{\Phi(y)} + \frac{y}{c^2} \bigg)
\end{aligned}$$
We know that $\lambda(y) > 0 \forall y$, so it reduces to show $\frac{\phi(y)}{\Phi(y)} + \frac{y}{c^2} >0$. When $y \geq 0$, it is true that $\frac{\phi(y)}{\Phi(y)} + \frac{y}{c^2}  \geq 0$. It remains to show $\frac{\phi(-y)}{\Phi(-y)} + \frac{-y}{c^2} > 0$ for $y >0$. However, this is a well-known property of Mill's ratio (see Fact 7.5.6 in \cite{tong}), which states that the upper bound of $\frac{\Phi(-y)}{\phi(-y)} $ is given by $\frac{y}{c^2}$ (recall that $y \sim N(0, c^2)$). We thus complete the proof. 

\end{proof}




\subsection{A4. Proof of Lemma 3.4}\label{a5.-proof-of-lemma-2.6}

\begin{proof}
Suppose $X_{ij}$ is ternary and $X_{ik}$ is continuous, then the sign expectation can break down as follows:
$$\begin{aligned}
&\quad \mathbb{E}[\text{sign}(X_{ij}- X_{i'j})(X_{ik}-X_{i'k})]\\
&=2\bigg(\mathbb{E}[I(X_{ij} = p, X_{ik} - X_{i'k} >0)] - \mathbb{E}[I(X_{i'j} = p, X_{ik} - X_{i'k} >0)]\bigg) \\
  &\quad + 2\bigg(\mathbb{E}\big[I(X_{ij} = p-1, X_{i'j} \leq p-2, X_{ik} - X_{i'k} >0)\big] - \mathbb{E}\big[I(X_{ij} \leq p-2,X_{i'j} = p-1, X_{ik} - X_{i'k} >0)\big]\bigg)\\
    &\quad + 2\bigg(\mathbb{E}\big[I(X_{ij} = p-2, X_{i'j} \leq p-3, X_{ik} - X_{i'k} >0)\big] - \mathbb{E}\big[I(X_{ij} \leq p-3,X_{i'j} = p-2, X_{ik} - X_{i'k} >0)\big]\bigg)\\
      &\quad ...\\
      &\quad + 2\bigg(\mathbb{E}\big[I(X_{ij} \leq 1 , X_{i'j}= 0 , X_{ik} - X_{i'k} >0)\big] - \mathbb{E}\big[I(X_{ij}= 0 ,X_{i'j} \leq 1, X_{ik} - X_{i'k} >0)\big]\bigg).
        \end{aligned}$$

However, it is a fact that 

$$\begin{aligned}
&\quad \mathbb{E}[I(X_{ij} = p, X_{ik} - X_{i'k} >0)] - \mathbb{E}[I(X_{i'j} = p, X_{ik} - X_{i'k} >0)]\\
&= \mathbb{P}(X_{i'j} = p, X_{ik} - X_{i'k} <0) - \mathbb{P}(X_{ij} = p, X_{ik} - X_{i'k} <0)\\
&= \big[1-\mathbb{P}(X_{i'j} \leq p-1, X_{ik} - X_{i'k} <0)\big] -  \big[1-\mathbb{P}(X_{ij} \leq p-1, X_{ik} - X_{i'k} < 0)\big]\\
&= \mathbb{P}(X_{ij} \leq p-1, X_{ik} - X_{i'k} < 0)-\mathbb{P}(X_{i'j} \leq p-1, X_{ik} - X_{i'k} <0)\\
&= \Phi_2(\Delta_j^{p},0,\sigma_{jk}/\sqrt{2}) - \Phi_2(\Delta_j^{p},0,-\sigma_{jk}/\sqrt{2})\\
&= 2\Phi_2(\Delta_j^{p},0,\sigma_{jk}/\sqrt{2}) - \Phi(\Delta_j^p)
\end{aligned}$$

and

$$\begin{aligned}
&\quad \mathbb{E}\big[I(X_{ij} = p-1, X_{i'j} \leq p-2, X_{ik} - X_{i'k} >0)\big] - \mathbb{E}\big[I(X_{ij} \leq p-2,X_{i'j} = p-1, X_{ik} - X_{i'k} >0)\big] \\
&=\mathbb{P}\big(X_{ij}\leq p-2, X_{i'j} = p-1,  X_{ik} - X_{i'k}<0\big)-\mathbb{P}\big(X_{ij}= p-1, X_{i'j} \leq p-2,  X_{ik} - X_{i'k}<0\big)\\
&= \bigg[\mathbb{P}\big(X_{ij}\leq p-2, X_{i'j} \leq p-1,  X_{ik} - X_{i'k}<0\big)-\mathbb{P}\big(X_{ij}\leq p-2, X_{i'j} \leq p-2,  X_{ik} - X_{i'k}<0\big)\bigg]\\
&\quad\quad -\bigg[\mathbb{P}\big(X_{ij}\leq p-1, X_{i'j} \leq p-2,  X_{ik} - X_{i'k}<0\big)-\mathbb{P}\big(X_{ij}\leq p-2, X_{i'j} \leq p-2,  X_{ik} - X_{i'k}<0\big)\bigg]\\
&=\mathbb{P}\big(X_{ij}\leq p-2, X_{i'j} \leq p-1,  X_{ik} - X_{i'k}<0\big)-\mathbb{P}\big(X_{ij}\leq p-1, X_{i'j} \leq p-2,  X_{ik} - X_{i'k}<0\big)\\
&=\Phi_3(\Delta_j^{p-1},\Delta_j^k,0) - \Phi_3(\Delta_j^{p},\Delta_j^{p-1},0)
\end{aligned}$$
and the other pairs of terms will follow the similar fashion.

Also notice that $(U_1 , U_2, \frac{V_1 - V_2}{\sqrt{2}}) \stackrel{d}{=} (U_2, U_1, -\frac{V_1 - V_2}{\sqrt{2}})$, so 
$$\begin{aligned}
&\quad \Phi_3(\Delta_j^{p-1},\Delta_j^p,0) +\Phi_3(\Delta_j^{p},\Delta_j^{p-1},0) \\
&= \mathbb{P}(U_1 < \Delta_j^{p-1}, U_2 < \Delta_j^p,  \frac{V_1 - V_2}{\sqrt{2}} < 0) +  \mathbb{P}(U_2 < \Delta_j^{p}, U_1 < \Delta_j^{p-1},  \frac{V_1 - V_2}{\sqrt{2}} > 0) \\
&= \mathbb{P}(U_1 < \Delta_j^{p-1}, U_2 < \Delta_j^p)\\
&= \Phi(\Delta_j^{p-1})\Phi(\Delta_j^{p}).
\end{aligned}$$
Therefore $$\Phi_3(\Delta_j^{p-1},\Delta_j^p,0) - \Phi_3(\Delta_j^{p},\Delta_j^{p-1},0) = 2\Phi_3(\Delta_j^{p-1},\Delta_j^p,0) - \Phi(\Delta_j^{p-1})\Phi(\Delta_j^{p}).$$

In addition, recall that $\hat{\Delta}_j^{p+1} = \Phi^{-1}(\frac{I(X_{ij} \leq p)}{n}) = \Phi^{-1}(1) = \infty$, so it holds that $$\Phi_2(\hat{\Delta}_j^p, 0, \sigma_{jk}/\sqrt{2}) = \Phi_3(\hat{\Delta}_j^p,\hat{\Delta}_j^{p+1},0),$$ so now we can alternatively express the bridge function as 

$$F(\sigma_{jk}; \mathbf{\Delta_j}) = \sum\limits_{l = 1}^p 4\Phi_3(\Delta_j^l, \Delta_j^{l+1},0) - 2\Phi(\Delta_j^l)\Phi(\Delta_j^{l+1}).$$

\end{proof}

\subsection{A5. Proof of Theorem 3.1}\label{a9.-proof-of-theorem-3.1}

We also need to show the convergence in probability for the correlation
estimator for ordinal-continuous mixed data:
\[ P(\sup ||\hat{r} - r ||  < C\sqrt{\log d}/n) > 1- d^{-1}. \]

In order to do so, we first show the Lipschitz continuity of the bridge
function where
\[ |F^{-1} ({\tau_1}; \mathbf{\Delta_j}) - F^{-1} ({\tau_{2}}; \mathbf{\Delta_j})| < L |\tau_1 - \tau_2| \]
for some constant \(L\), and
\(\mathbf{\Delta_j} = (\Delta_j^1, \Delta_j^2)\), which is equivalent to
show that
\[\frac{\partial F(r; \mathbf{\Delta_j})}{\partial r} > \frac{1}{L}. \]
Recall that
\[\frac{\partial}{\partial r}  4\Phi_2(\Delta_j^{k},0,r/\sqrt{2}) - 2\Phi(\Delta_j^{k}) > \frac{1}{L_1}\]
from \cite{Fanetal2017}, so we are left with proving
\[\frac{\partial}{\partial r}  \Phi_3(\Delta_j^{k-1},\Delta_j^k,0)  > \frac{1}{L}\]
for some constant \(L\).

However, since we just showed that
\[\dfrac{\partial \Phi_3(\Delta_j^{k-1},\Delta_j^k,0)}{\partial r}  = \frac{1}{\sqrt{2}}\int_{-\infty }^{\Delta_j^k}   \phi_3(\Delta_{j}^{k-1},x_2,\Delta_j^k)dx_2,\]
it is immediate that
\[\dfrac{\partial \Phi_3(\Delta_j^{k-1},\Delta_j^k,0)}{\partial r} > \frac{1}{\sqrt{2}}\int_{-\infty }^{\Delta_j^k - \epsilon}   \phi_3(\Delta_{j}^{k-1},x_2,\Delta_j^k)dx_2\]
for any positive \(\epsilon\).

Next we will need to prove the upper bound using Hoeffding's inequality.

\begin{proof}
By Lipschitz continuity of $\Phi^{-1}(\cdot)$ in $[\Phi(-2M),\Phi(2M)]$, we know that under the event $A_{j,1} = \{|\hat{\Delta}_j^1| \leq 2M \}$, there exists a Lipschitz constant $L_1$ such that
$$\begin{aligned}
|\hat{\Delta}_j^1 - {\Delta}_j^1| &= \bigg|\Phi^{-1}\bigg( \frac{\sum_{i = 1}^n I(X_{ij} = 0)}{n} \bigg) - \Phi^{-1}(\Phi(\Delta_j^1))\bigg| \\
&\leq L_1 \bigg|\frac{\sum_{i = 1}^n I(X_{ij} = 0)}{n} - \Phi(\Delta_j^1)\bigg|.
\end{aligned}$$
The exception probability is controlled by
$$\begin{aligned}
P(A_{j,1}^c) &= P(|\hat{\Delta}_j^1| > 2M) \\
&\leq P(|\hat{\Delta}_j^1| - |{\Delta}_j^1| > M) \\
&\leq P(|\hat{\Delta}_j^1 -{\Delta}_j^1| > M)\\
&\leq P\bigg( \bigg|\frac{\sum_{i = 1}^n I(X_{ij} = 0)}{n} - \Phi(\Delta_j^1)\bigg| > \frac{M}{L_1} \bigg)\\
&\leq 2\exp\bigg(-\frac{2M^2 n}{L_1^2} \bigg) \quad\quad\quad\quad\quad\quad\quad\quad\quad\quad\quad\quad\quad\quad\quad \text{(by Hoeffding's inequality).}
\end{aligned}$$

Likewise, under $A_{j,2} = \{|\hat{\Delta}_j^2| \leq 2M \}$, we have 
$$|\hat{\Delta}_j^2 - {\Delta}_j^2| \leq L_1 \bigg|\frac{\sum_{i = 1}^n I(X_{ij} \leq 1)}{n} - \Phi(\Delta_j^2)\bigg|;$$
and $$P(A_{j,2}^c) \leq 2\exp\bigg(-\frac{2M^2 n}{L_1^2} \bigg).$$
Now we define the event $A_j = \bigcap\limits_{l = 1}^2 A_{j,l}$, as a result we have
$$\begin{aligned}
P(A_j^c) &= P(\bigcup\limits_{l = 1}^2 A_{j,l}^c) \\
&\leq \sum_{l = 1}^2 P(A_{j,l}^c)\\
&\leq 4\exp\bigg(-\frac{2M^2 n}{L_1^2} \bigg).
\end{aligned}$$

For any $t>0$, we have
$$\begin{aligned}
&\quad P(|F^{-1} (\hat{\tau}^a; \hat{\mathbf{\Delta_j}}) - r | \geq t)\\
&= P(\big\{|F^{-1} (\hat{\tau}^a; \hat{\mathbf{\Delta_j}}) - r | \geq t\big\} \cap A_j) + P(\big\{|F^{-1} (\hat{\tau}^a; \hat{\mathbf{\Delta_j}}) - r | \geq t\big\} \cap A_j^c)\\
&\leq P(\big\{|F^{-1} (\hat{\tau}^a; \hat{\mathbf{\Delta_j}}) - r | \geq t\big\} \cap A_j) + P(A_j^c).
\end{aligned}$$

Recall that $F^{-1}(\tau^a;\mathbf{\Delta})$ is Lipschitz continuous on $[-1,1]$ with Lipschitz constant $L_4$, we then have
$$\begin{aligned}
&\quad P(\big\{|F^{-1} (\hat{\tau}^a; \hat{\mathbf{\Delta_j}}) - r | \geq t\big\} \cap A_j) \\
&= P(\big\{|F^{-1} (\hat{\tau}^a; \hat{\mathbf{\Delta_j}}) - F^{-1}(F(r;\hat{\mathbf{\Delta_j}} );\hat{\mathbf{\Delta_j}}) | \geq t\big\} \cap A_j) \\
&\leq P(\{ L_4 | \hat{\tau}^a - F(r; \hat{\mathbf{\Delta_j}}) | >t \} \cap A_j) \\
&\leq P(\{L_4 | \hat{\tau}^a - F(r; {\mathbf{\Delta_j}}) | + L_4 | F(r; {\mathbf{\Delta_j} })- F(r; \hat{\mathbf{\Delta_j}}) | > t\} \cap A_j)\\
&\leq P(\{L_4 | \hat{\tau}^a - F(r; {\mathbf{\Delta_j}}) | > \frac{t}{2}\} \cap A_j) + P(\{L_4 | F(r; {\mathbf{\Delta_j} }) - F(r; \hat{\mathbf{\Delta_j}}) | > \frac{t}{2}\} \cap A_j)\\
&\leq P(L_4 | \hat{\tau}^a - F(r; {\mathbf{\Delta_j}}) | > \frac{t}{2}) +  P(\{L_4 | F(r; {\mathbf{\Delta_j} }) - F(r; \hat{\mathbf{\Delta_j}}) | > \frac{t}{2}\} \cap A_j)\\
&\equiv I_1 + I_2.
\end{aligned}$$

Since $\hat{\tau}^a$ is a U-statistic with bounded kernel, it is immediate by Hoeffding's inequality that $$I_1 = P(L_4 | \hat{\tau}^a - F(r; {\mathbf{\Delta_j}}) | > \frac{t}{2}) \leq 2\exp\big(-\frac{nt^2}{2L_4^2} \big).$$

Let $\Phi_{21} (x,y,t) = \frac{\partial \Phi_2(x,y,t)}{\partial x}$, $\Phi_{31}(x,y,z) =  \frac{\partial \Phi_3(x,y,t)}{\partial x}$, and $\Phi_{32}(x,y,z) =  \frac{\partial \Phi_3(x,y,t)}{\partial y}$. For $I_2$, we have  

$$\begin{aligned}
&\quad | F(r; {\mathbf{\Delta_j} }) - F(r; \hat{\mathbf{\Delta_j}}) | \\ 
&\leq 4 |\Phi_2(\Delta_j^2, 0, r/\sqrt{2}) - \Phi_2(\hat{\Delta}_j^2, 0, r/\sqrt{2}) | + 2 | \Phi(\Delta_j^2) - \Phi(\hat{\Delta}_j^2)  | \\
&\quad \quad + 4 |\Phi_3(\Delta_j^1, \Delta_j^2, 0) - \Phi_3(\hat{\Delta}_j^1, \hat{\Delta}_j^2, 0) |  + 2 |\Phi(\Delta_j^1)\Phi(\Delta_j^2) - \Phi(\hat{\Delta}_j^1)\Phi(\hat{\Delta}_j^2)| \\
&\leq 4\Phi_{21} (\zeta_1) |\Delta_j^2 - \hat{\Delta}_j^2 |  + 2\phi(\zeta_2) |\Delta_j^2 - \hat{\Delta}_j^2 | + 4\Phi_{31}(\zeta_3) |\Delta_j^1 - \hat{\Delta}_j^1 | + 4 \Phi_{32}(\zeta_4) |\Delta_j^2 - \hat{\Delta}_j^2| \\
&\quad\quad 2\Phi(\hat{\Delta}_j^1)\phi(\zeta_5) |\Delta_j^2 - \hat{\Delta}_j^2| + 2\Phi(\hat{\Delta}_j^2)\phi(\zeta_6) |\Delta_j^1 - \hat{\Delta}_j^1|.
\end{aligned}$$
It has been shown that $\Phi_{21}(x,y,t) \leq \frac{1}{\sqrt{2\pi}}$ from \cite{Fanetal2017} . For the upper bound of $\Phi_{31}(x,y,z)$, we know that the conditional distribution of $(Y,Z)$ given $X$ is bivariate normal:
$$ Y, Z | X=x \sim N\bigg(\begin{bmatrix}
0 \\ 
\frac{xr}{\sqrt{2}}
\end{bmatrix}, \begin{bmatrix}  
1 & -r/\sqrt{2} \\
-r/\sqrt{2}  & 1 
\end{bmatrix} \bigg).$$ 
Let $\phi_2(y,z | x)$ denote the density function for the conditional distribution, and $\Phi_2(y, z | x)$ denote the distribution function.
Therefore 
$$\Phi_3(\Delta_j^1,\Delta_j^2,0) = \int_{-\infty}^{\Delta_j^1} \int_{-\infty}^{\Delta_j^2} \int_{-\infty}^0 \phi_2(y, z | x) \phi(x) dzdydx = \int_{-\infty}^{\Delta_j^1} \Phi_2 (\Delta_j^2, 0 | x) \phi(x)dx$$ hence $$\Phi_{31} = \frac{\partial \Phi_3(\Delta_j^1,\Delta_j^2,0)}{\partial \Delta_j^1} = \frac{\partial}{\partial \Delta_j^1} \int_{-\infty}^{\Delta_j^1} \Phi_2 (\Delta_j^2, 0 | x) \phi(x)dx = \Phi_2(\Delta_j^2, 0 | \Delta_j^1)\phi(\Delta_j^1) \leq \frac{1}{\sqrt{2\pi}}$$ and 
$$| F(r; {\mathbf{\Delta_j} }) - F(r; \hat{\mathbf{\Delta_j}}) | \leq \frac{12}{\sqrt{2\pi}} |\Delta_j^2 - \hat{\Delta}_j^2| + \frac{6}{\sqrt{2\pi}} |\Delta_j^1 - \hat{\Delta}_j^1|.$$

As a result, the upper bound for $I_2$ is established:
$$\begin{aligned}
I_2 &\leq P(\{\frac{12}{\sqrt{2\pi}} L_4 |\Delta_j^2 - \hat{\Delta}_j^2| + \frac{6}{\sqrt{2\pi}} L_4|\Delta_j^1 - \hat{\Delta}_j^1| > \frac{t}{2} \}  \cap A_j) \\
&\leq P(\bigg|\frac{\sum_{i = 1}^n I(X_{ij} \leq 1)}{n} - \Phi(\Delta_j^2)\bigg| > \frac{t\sqrt{2\pi}}{48L_1 L_4}) + P(\bigg|\frac{\sum_{i = 1}^n I(X_{ij} = 0)}{n} - \Phi(\Delta_j^1)\bigg| > \frac{t\sqrt{2\pi}}{24 L_1 L_4}) \\
&\leq 2\exp( - \frac{nt^2\pi}{48^2 L_1^2 L_4^2 }) + 2\exp( - \frac{nt^2\pi}{24^2 L_1^2 L_4^2 }).
\end{aligned}$$

So putting together we have 
$$P \big(\big| \hat{r} - r \big| > t \big) \leq 4\exp\bigg(-\frac{2M^2 n}{L_1^2} \bigg) + 2\exp\big(-\frac{nt^2}{2L_4^2} \big) + 2\exp( - \frac{nt^2\pi}{48^2 L_1^2 L_4^2 }) + 2\exp( - \frac{nt^2\pi}{24^2 L_1^2 L_4^2 }).$$

\end{proof}

\subsection{A6. Proof of Theorem 3.2}\label{a10.-proof-of-theorem-3.2}

In this section, we show that the correlation estimator for \(p\)-level
mixed data also converge to the true correlation parameter in
probability, namely
\[ P(\sup ||\hat{r} - r ||  < C\sqrt{\log d}/n) > 1- d^{-1}. \]

\begin{proof}
We begin the proof by showing the Lipschitz continuity of the bridge function. Recall that 
$$\frac{\partial F(r; \mathbf{\Delta_j})}{\partial r} = 4\sum_{l = 1}^k \frac{1}{\sqrt{2}} \int_{-\infty}^{\Delta_j^{l+1}}\phi_3(\Delta_j^l, x, 0)dx,$$
let $\delta_{m} = \min\limits_{l = 1, ... , k} \quad 4\frac{1}{\sqrt{2}} \int_{-\infty}^{\Delta_j^{l+1}}\phi_3(\Delta_j^l, x, 0)dx$ then we have the Lipschitz constant $L$ for $F^{-1} (\tau^a; \mathbf{\Delta_j})$ such that
$$\frac{\partial F(r; \mathbf{\Delta_j})}{\partial r}  \geq \dfrac{1}{L} = K\delta_m.$$

Consequently, for $\mathbf{\hat{\Delta}_j} \in A_j$, the Lipschitz continuity of $F^{-1}(\tau^a; \mathbf{\Delta_j})$ gives rise to
$$|F^{-1} (\hat{\tau}^a; \hat{\mathbf{\Delta_j}}) - F^{-1}(F(r;\hat{\mathbf{\Delta_j}} );\hat{\mathbf{\Delta_j}}) |  \leq L | \hat{\tau}^a - F(r; \hat{\mathbf{\Delta_j}}) |.$$

Also recall that $\Phi^{-1}(\cdot)$ is Lipschitz continuous in $[\Phi(-2M),\Phi(2M)]$, we have 
a Lipschitz constant $L_1$ such that
$$\begin{aligned}
|\hat{\Delta}_j^1 - {\Delta}_j^1| &= \bigg|\Phi^{-1}\bigg( \frac{\sum_{i = 1}^n I(X_{ij} = 0)}{n} \bigg) - \Phi^{-1}(\Phi(\Delta_j^1))\bigg| \\
&\leq L_1 \bigg|\frac{\sum_{i = 1}^n I(X_{ij} = 0)}{n} - \Phi(\Delta_j^1)\bigg|.
\end{aligned}$$
The exception probability is controlled by
$$\begin{aligned}
P(A_{j,1}^c) &= P(|\hat{\Delta}_j^1| > 2M) \\
&\leq P(|\hat{\Delta}_j^1| - |{\Delta}_j^1| > M) \\
&\leq P(|\hat{\Delta}_j^1 -{\Delta}_j^1| > M)\\
&\leq P\bigg( \bigg|\frac{\sum_{i = 1}^n I(X_{ij} = 0)}{n} - \Phi(\Delta_j^1)\bigg| > \frac{M}{L_1} \bigg)\\
&\leq 2\exp\bigg(-\frac{2M^2 n}{L_1^2} \bigg) \quad\quad\quad\quad\quad\quad\quad\quad\quad\quad\quad\quad\quad\quad\quad \text{(by Hoeffding's inequality).}
\end{aligned}$$

Likewise, under $A_{j,l} = \{|\hat{\Delta}_j^l| \leq 2M \}$, we have 
$$|\hat{\Delta}_j^l - {\Delta}_j^l| \leq L_1 \bigg|\frac{\sum_{i = 1}^n I(X_{ij} \leq l - 1)}{n} - \Phi(\Delta_j^l)\bigg|;$$
and $$P(A_{j,l}^c) \leq 2\exp\bigg(-\frac{2M^2 n}{L_1^2} \bigg).$$ 
Now we define the event $A_j = \bigcap\limits_{l = 1}^2 A_{j,l}$, as a result we have
$$\begin{aligned}
P(A_j^c) &= P(\bigcup\limits_{l = 1}^{k-1} A_{j,l}^c) \\
&\leq \sum_{l = 1}^{k-1} P(A_{j,l}^c)\\
&\leq 2(k-1)\exp\bigg(-\frac{2M^2 n}{L_1^2} \bigg).
\end{aligned}$$

For any $t>0$, we have
$$\begin{aligned}
&\quad P(|F^{-1} (\hat{\tau}^a; \hat{\mathbf{\Delta_j}}) - r | \geq t)\\
&= P(\big\{|F^{-1} (\hat{\tau}^a; \hat{\mathbf{\Delta_j}}) - r | \geq t\big\} \cap A_j) + P(\big\{|F^{-1} (\hat{\tau}^a; \hat{\mathbf{\Delta_j}}) - r | \geq t\big\} \cap A_j^c)\\
&\leq P(\big\{|F^{-1} (\hat{\tau}^a; \hat{\mathbf{\Delta_j}}) - r | \geq t\big\} \cap A_j) + P(A_j^c).
\end{aligned}$$

Recall that $F^{-1}(\tau^a;\mathbf{\Delta})$ is Lipschitz continuous on $[-1,1]$ with Lipschitz constant $L$, we then have
$$\begin{aligned}
&\quad P(\big\{|F^{-1} (\hat{\tau}^a; \hat{\mathbf{\Delta_j}}) - r | \geq t\big\} \cap A_j) \\
&= P(\big\{|F^{-1} (\hat{\tau}^a; \hat{\mathbf{\Delta_j}}) - F^{-1}(F(r;\hat{\mathbf{\Delta_j}} );\hat{\mathbf{\Delta_j}}) | \geq t\big\} \cap A_j) \\
&\leq P(\{ L | \hat{\tau}^a - F(r; \hat{\mathbf{\Delta_j}}) | >t \} \cap A_j) \\
&\leq P(\{L | \hat{\tau}^a - F(r; {\mathbf{\Delta_j}}) | + L | F(r; {\mathbf{\Delta_j} })- F(r; \hat{\mathbf{\Delta_j}}) | > t\} \cap A_j)\\
&\leq P(\{L | \hat{\tau}^a - F(r; {\mathbf{\Delta_j}}) | > \frac{t}{2}\} \cap A_j) + P(\{L | F(r; {\mathbf{\Delta_j} }) - F(r; \hat{\mathbf{\Delta_j}}) | > \frac{t}{2}\} \cap A_j)\\
&\leq P(L | \hat{\tau}^a - F(r; {\mathbf{\Delta_j}}) | > \frac{t}{2}) +  P(\{L | F(r; {\mathbf{\Delta_j} }) - F(r; \hat{\mathbf{\Delta_j}}) | > \frac{t}{2}\} \cap A_j)\\
&\equiv I_1 + I_2.
\end{aligned}$$

Since $\hat{\tau}^a$ is a U-statistic with bounded kernel, it is immediate by Hoeffding's inequality that $$I_1 = P(L | \hat{\tau}^a - F(r; {\mathbf{\Delta_j}}) | > \frac{t}{2}) \leq 2\exp\big(-\frac{nt^2}{2L^2} \big) = 2\exp\bigg(- \dfrac{nt^2k^2\delta_m^2}{2} \bigg).$$

Let $\Phi_{21} (x,y,t) = \frac{\partial \Phi_2(x,y,t)}{\partial x}$, $\Phi_{31}(x,y,z) =  \frac{\partial \Phi_3(x,y,t)}{\partial x}$, and $\Phi_{32}(x,y,z) =  \frac{\partial \Phi_3(x,y,t)}{\partial y}$. For $I_2$, we have  

$$\begin{aligned}
&\quad | F(r; {\mathbf{\Delta_j} }) - F(r; \hat{\mathbf{\Delta_j}}) | \\ 
&= \bigg|\sum\limits_{l = 1}^k 4\Phi_3(\Delta_j^l, \Delta_j^{l+1}, 0 ) - 2\Phi(\Delta_j^l)\Phi(\Delta_j^{l+1}) - 4\Phi_3(\hat{\Delta}_j^l, \hat{\Delta}_j^{l+1}, 0 )+ 2\Phi(\hat{\Delta}_j^l)\Phi(\hat{\Delta}_j^{l+1})\bigg| \\
&\leq \sum\limits_{l = 1}^k 4 \bigg|\Phi_3(\Delta_j^l, \Delta_j^{l+1}, 0 ) -\Phi_3(\hat{\Delta}_j^l, \hat{\Delta}_j^{l+1}, 0 )\bigg| + 2\bigg| \Phi(\Delta_j^l)\Phi(\Delta_j^{l+1})-\Phi(\hat{\Delta}_j^l)\Phi(\hat{\Delta}_j^{l+1}) \bigg| \\
&\leq  4 \sum\limits_{l = 1}^k \bigg(\Phi_{31}(\zeta_{1,l}) |\Delta_j^l - \hat{\Delta}_j^l | + \Phi_{32}(\zeta_{2,l}) |\Delta_{j}^{l+1} - \hat{\Delta}_{j}^{l+1} | \bigg) \\
&\quad\quad\quad +2 \sum\limits_{l = 1}^k \bigg( \Phi(\hat{\Delta}_j^l) \phi(\eta_{1,l})|\Delta_{j}^{l+1} - \hat{\Delta}_{j}^{l+1} | +  \Phi(\hat{\Delta}_j^{l+1}) \phi(\eta_{2,l})|\Delta_{j}^{l} - \hat{\Delta}_{j}^{l} |\bigg) \\
&\leq 4 \sum\limits_{l = 1}^{k} \frac{1}{\sqrt{2\pi}}|\Delta_{j}^{l} - \hat{\Delta}_{j}^{l} | + \frac{1}{\sqrt{2\pi}}|\Delta_{j}^{l+1} - \hat{\Delta}_{j}^{l+1} | \\
&\quad\quad\quad +2 \sum\limits_{l = 1}^k \frac{1}{\sqrt{2\pi}}|\Delta_{j}^{l} - \hat{\Delta}_{j}^{l} | + \frac{1}{\sqrt{2\pi}}|\Delta_{j}^{l+1} - \hat{\Delta}_{j}^{l+1} | \\
&= 6\sum\limits_{l = 2}^{k-1} \frac{\sqrt{2}}{\sqrt{\pi}}|\Delta_{j}^{l} - \hat{\Delta}_{j}^{l} | + \frac{6}{\sqrt{2\pi}} |\Delta_{j}^{1} - \hat{\Delta}_{j}^{1} | + \frac{6}{\sqrt{2\pi}} |\Delta_{j}^{k} - \hat{\Delta}_{j}^{k} |.
\end{aligned}$$

We now can establish the bound for $I_2$:

$$\begin{aligned}
I_2 &\leq P(\{6\sum\limits_{l = 2}^{k-1} \frac{\sqrt{2}}{\sqrt{\pi}}|\Delta_{j}^{l} - \hat{\Delta}_{j}^{l} | + \frac{6}{\sqrt{2\pi}} |\Delta_{j}^{1} - \hat{\Delta}_{j}^{1} | + \frac{6}{\sqrt{2\pi}} |\Delta_{j}^{k} - \hat{\Delta}_{j}^{k} |> \frac{t}{2L} \}  \cap A_j) \\
&\leq P(\bigg|\frac{\sum_{i = 1}^n I(X_{ij} = 0)}{n} - \Phi(\Delta_j^1)\bigg| > \frac{t\sqrt{2\pi}}{12 L_1 Lk})\\
&\quad\quad + P(\bigg|\frac{\sum_{i = 1}^n I(X_{ij} \leq k-1)}{n} - \Phi(\Delta_j^k) \bigg| > \frac{t\sqrt{2\pi}}{12 L_1 Lk}) \\
&\quad\quad + \sum\limits_{l=2}^{k-1} P(\bigg|\frac{\sum_{i = 1}^n I(X_{ij} \leq l-1)}{n} - \Phi(\Delta_j^l) \bigg| > \frac{t\sqrt{2\pi}}{24 L_1 L k})\\
&\leq 2\exp( - \frac{4nt^2\pi}{12^2 L_1^2 L^2 k^2 })+2\exp( - \frac{4nt^2\pi}{12^2 L_1^2 L^2 k^2}) + 2(k-2)\exp( - \frac{4nt^2\pi}{24^2 L_1^2 L^2 k^2})\\
&= 4\exp(- \frac{4nt^2\pi}{12^2 L_1^2  }\delta_m^2)+2(k-2)\exp(- \frac{4nt^2\pi}{24^2 L_1^2  }\delta_m^2)\\
&\leq 2k\exp(- \frac{4nt^2\pi}{24^2 L_1^2  }\delta_m^2).
\end{aligned}$$

So putting everything together we have

$$\begin{aligned}
P(||\hat{r} - r || >t) &\leq 2(k-1)\exp\bigg(-\frac{2M^2 n}{L_1^2} \bigg) +  2\exp\bigg(- \dfrac{nt^2k^2\delta_m^2}{2} \bigg) \\
&\quad+2k\exp(- \frac{4nt^2\pi}{24^2 L_1^2  }\delta_m^2)
\end{aligned}$$
implying that 
$$\begin{aligned}
P(\sup ||\hat{r} - r || >t) &\leq \sum_{j,k} P(||\hat{r}_{jk} - r_{jk}  || >t) \\
&\leq 2d^2(k-1)\exp\bigg(-\frac{2M^2 n}{L_1^2} \bigg) +  2d^2\exp\bigg(- \dfrac{nt^2k^2\delta_m^2}{2} \bigg) \\
&\quad+2d^2k\exp(- \frac{4nt^2\pi}{24^2 L_1^2  }\delta_m^2).            
\end{aligned}$$

Therefore at fixed $k$, taking $t = C\sqrt{\frac{\log (dk)}{n}}$ we have
$$P(\sup ||\hat{r} - r ||  < C\sqrt{\frac{\log dk}{n}}) > 1- d^{-1}. $$
\end{proof}

\subsection{A7. Proof of Lemma 4.1}\label{a6.-proof-of-lemma-2.7}

\begin{proof}
The 1st-order Taylor expansion gives rise to 

$$\begin{aligned}
\mathbb{E}(\hat{\tau}_{jk}^b) &= \mathbb{E}\bigg[\frac{C-D}{\sqrt{\big[\binom{n}{2} - t_{X_j}\big] \big[\binom{n}{2}- t_{X_k}\big]}}\bigg] \\
&\approx \dfrac{\mathbb{E}(C-D)}{\mathbb{E}\big(\sqrt{\big[\binom{n}{2} - t_{X_j}\big] \big[\binom{n}{2}- t_{X_k}\big]}\big)}\\
&= \dfrac{2\big[\Phi_2(\Delta_j, \Delta_k, \sigma_{jk}) - \Phi(\Delta_j)\Phi(\Delta_k) \big]}{\sqrt{1-p_j}\sqrt{1-p_k}}
\end{aligned}$$
where $p_j$ is the probability of getting a tied pair at $X_j$, and likewise for $p_k$. 

We know that 

$$\begin{aligned}
1- p_j &= P([(1,x_{ik})(0,x_{i'k})]) + P([(0,x_{ik})(1,x_{i'k})]) \\
&= 2P([(1,x_{ik})(0,x_{i'k})]) \\
&= 2\bigg[\Phi(\Delta_j)\big(1-\Phi(\Delta_j)\big) \bigg]
\end{aligned}$$

and likewise $1- p_k = 2\bigg[\Phi(\Delta_k)\big(1-\Phi(\Delta_k)\big) \bigg].$
Combining these results, we have 
$$F_b (\sigma_{jk};\Delta_j,\Delta_k) = \dfrac{\Phi_2(\Delta_j,\Delta_k,\sigma_{jk}) -\Phi(\Delta_j) \Phi(\Delta_k) }{\sqrt{(\Phi(\Delta_j)- \Phi(\Delta_j)^2)(\Phi(\Delta_k)- \Phi(\Delta_k)^2)}}.$$

\end{proof}

\subsection{A8. Proof of Lemma 4.2}\label{a7.-proof-of-lemma-2.8}

\begin{proof}
Since $\mathbf{X}_{k}$ is continuous, we do not need to consider tieing at $\mathbf{X}_k$. Therefore, the bridge function is easily derived as

$$\begin{aligned}
\mathbb{E}(\hat{\tau}_{jk}^b) &= \mathbb{E}\bigg[\frac{C-D}{\sqrt{\big[\binom{n}{2} - t_{X_j}\big] \big[\binom{n}{2}\big]}}\bigg] \\
&\approx \dfrac{\mathbb{E}(C-D)}{\mathbb{E}\big(\sqrt{\big[\binom{n}{2} - t_{X_j}\big] \big[\binom{n}{2}\big]}\big)}\\
&= \dfrac{4\Phi_2(\Delta_j, 0, \sigma_{jk}/\sqrt{2})-2\Phi(\Delta_j)}{\sqrt{1-p_j}}\\
&= \dfrac{4\Phi_2(\Delta_j, 0, \sigma_{jk}/\sqrt{2})-2\Phi(\Delta_j)}{\sqrt{2(\Phi(\Delta_j))- 2(\Phi(\Delta_j))^2}}
\end{aligned}$$
where in the second last step we adopt the result from Kendall's $\tau^a$ version bridge function in \cite{Fanetal2017} and the last step uses the result derived in A7.

\end{proof}

\subsection{A9. Proof of Lemma 4.3}\label{a8.-proof-of-lemma-2.9}

\begin{proof}

2nd order Taylor expansion gives:
$$\begin{aligned}
E(Y/X) &\approx \frac{\mu_Y}{\mu_X} + \sigma_X^2 \frac{\mu_Y}{\mu_X^3} - \frac{\sigma_{XY}}{\mu_X^2}\\
&= \frac{\mu_Y}{\mu_X} + \frac{1}{\mu_X^2}\bigg(\sigma_X^2\frac{\mu_Y}{\mu_X} - \rho \sigma_X \sigma_Y \bigg).
\end{aligned}$$

Therefore we have 

$$\begin{aligned}
E(\hat{\tau}_{jk}^b) &= E(\dfrac{\sum\limits_{1\leq i < i' \leq n} (X_{ij}-X_{i'j})\text{sign}(X_{ik}-X_{i'k})}{\sqrt{\binom{n}{2}- \sum_i \binom{n_{i+}}{2}} \sqrt{\binom{n}{2}}})\\
&= E(\frac{\sqrt{\binom{n}{2}}\hat{\tau}_{jk}^a}{\sqrt{\binom{n}{2}-T}})\\
&\approx \frac{\sqrt{\binom{n}{2}}E[\hat{\tau}_{jk}^a]}{E[\sqrt{\binom{n}{2}-T}]}+... \\
&\quad \quad\quad\quad...   \frac{1}{\bigg[E[\sqrt{\binom{n}{2}-T}]\bigg]^2}\bigg[\binom{n}{2}\text{var}\bigg(\sqrt{\binom{n}{2}-T}\bigg)\frac{\sqrt{\binom{n}{2}}E[\hat{\tau}_{jk}^a]}{E[\sqrt{\binom{n}{2}-T}]} - \text{cov}\big(\sqrt{\binom{n}{2}}\hat{\tau}_{jk}^a,\sqrt{\binom{n}{2}-T}\big)\bigg].
\end{aligned}$$

We compute each part separately. First, note that $E[\hat{\tau}_{jk}^a]$ can be directly adopted from \cite{Fanetal2017}, namely
$$\begin{aligned}
E[\hat{\tau}_{jk}^a] &= E\big[\sum\limits_{1\leq i < i' \leq n} (X_{ij}-X_{i'j})\text{sign}(X_{ik}-X_{i'k})\big]\\
&= 4\Phi_2(\Delta_j, 0, r/\sqrt{2})-2\Phi(\Delta_j).\end{aligned}$$

For $E\bigg[{\sqrt{\binom{n}{2}-t_{X_j}}}\bigg]$, we know that the number of ties are $\binom{n_0}{2} + \binom{n_1}{2}$ where $n_0$ is the number of $X_{ij} = 0$ for $i = 1,\ldots, n$ and $n_1$ is the number of $X_{ij} = 1$ for $i = 1, \ldots, n$. Also recall that $\mathbb{P}(X_{ij} = 0) = \Phi(\Delta_j)$, therefore we have 
$$\begin{aligned}
E\bigg[{\sqrt{\binom{n}{2}-T}}\bigg] &= E\bigg[\sqrt{\binom{n}{2} - \binom{n_0}{2} - \binom{n_1}{2}}\bigg]\\
&=\sum_{n_0=0}^n \bigg[\sqrt{\binom{n}{2} - \binom{n_0}{2} - \binom{n-n_0}{2}}\bigg] \binom{n}{n_0}\big(\Phi(\Delta_j)\big)^{n_0}\big(1-\Phi(\Delta_j)\big)^{n-n_0}
\end{aligned}$$
and consequently
$$\begin{aligned}
\text{var}\bigg[{\sqrt{\binom{n}{2}-T}}\bigg] &= E\bigg[{{\binom{n}{2}-T}}\bigg] - (E\bigg[{\sqrt{\binom{n}{2}-T}}\bigg])^2\\
&= \binom{n}{2}\big(2\Phi(\Delta_j) - 2[\Phi(\Delta_j)]^2 \big) - (E\bigg[{\sqrt{\binom{n}{2}-T}}\bigg])^2.
\end{aligned}$$

As for $\text{cov}\bigg(\sqrt{\binom{n}{2}}\tau_{jk}^a,\sqrt{\binom{n}{2}-T}\bigg)$, we know that $\tau_{jk}^a = \frac{C-D}{\binom{n}{2}}$, and $\binom{n}{2} -T= C+D$, so we can instead compute
$$\begin{aligned}
\text{cov}\bigg((C-D),\sqrt{C+D}\bigg) &=E[(C-D)\sqrt{C+D}]-E(C-D)E[\sqrt{C+D}] \\
&=E[(C-D)\sqrt{C+D}]-E(C-D)E[\sqrt{C+D}]
\end{aligned}$$

where we can compute $E(\sqrt{(C-D)(C^2-D^2)})$ from the fact that $(C,D)$ follows a multinomial distribution with parameters
$$\begin{aligned}
p_C &= 2\mathbb{P}[(X_{ij} = 0, X_{i'j} =1, (X_{ik} - X_{i'k})/\sqrt{2} <0)]\\
&= 2[\Phi_2(\Delta_j, 0, \sigma_{jk}/\sqrt{2}) - \Phi_3(\Delta_j, \Delta_j, 0)]
\end{aligned}$$
and

$$\begin{aligned}
p_D &= 2\mathbb{P}[(X_{ij} = 1, X_{i'j} = 0, (X_{ik} - X_{i'k})/\sqrt{2} <0)]\\
&= 2[\Phi_2(\Delta_j, 0, -\sigma_{jk}/\sqrt{2}) - \Phi_3(\Delta_j, \Delta_j, 0)]
\end{aligned}$$

so $$E[(C-D)\sqrt{C+D}] = \sum_{(C,D) \in S} (C-D)\sqrt{C+D}\frac{\binom{n}{2}}{C!D!\bigg(\binom{n}{2}-C-D \bigg)!} p_C^C p_D^D (1-p_C-p_D)^{\binom{n}{2}-C-D }$$

with the sample space of $(C,D)$ being $S = \{(C,D) :  C\in \mathbb{Z}^+, D \in \mathbb{Z}^+, C+D \leq n \}$.

Putting these together, we have 
$$\begin{aligned} \text{cov}\bigg(\sqrt{\binom{n}{2}}\tau_{jk}^a,\sqrt{\binom{n}{2}-t_{X_j}}\bigg) &= \sum_{(C,D) \in S}\bigg\{ (C-D)\sqrt{(C+D)}\frac{\sqrt{\binom{n}{2}}}{C!D!\bigg(\binom{n}{2}-C-D \bigg)!}\cdot\\
&\quad\quad\quad\quad p_C^C p_D^D (1-p_C-p_D)^{\binom{n}{2}-C-D}\bigg\}- \sqrt{\binom{n}{2}}\mathbb{E}(\hat{\tau}_{jk}^a)\mathbb{E}\big[{\sqrt{\binom{n}{2} - t_{X_j}}}\big].
\end{aligned}$$

\end{proof}

\newpage
\bibliographystyle{chicago}  
\bibliography{bibref} 

\begin{thebibliography}{}

\bibitem[\protect\citeauthoryear{Adler, Falk, Friedler, Nix, Rybeck,
  Scheidegger, Smith, and Venkatasubramanian}{Adler
  et~al.}{2018}]{adler2018auditing}
Adler, P., C.~Falk, S.~A. Friedler, T.~Nix, G.~Rybeck, C.~Scheidegger,
  B.~Smith, and S.~Venkatasubramanian (2018).
\newblock Auditing black-box models for indirect influence.
\newblock {\em Knowledge and Information Systems\/}~{\em 54\/}(1), 95--122.

\bibitem[\protect\citeauthoryear{Agresti}{Agresti}{2010}]{Agresti2010}
Agresti, A. (2010).
\newblock {\em Analysis of Ordinal Categorical Data}.
\newblock John Wiley \& Sons, Inc.

\bibitem[\protect\citeauthoryear{Andrews}{Andrews}{1985}]{Andrews}
Andrews, D.~F. (1985).
\newblock {\em Data : A Collection of Problems from Many Fields for the Student
  and Research Worker}.
\newblock Springer, New York.

\bibitem[\protect\citeauthoryear{Angwin, Larson, Mattu, and Kirchner}{Angwin
  et~al.}{2016}]{angwin2016machine}
Angwin, J., J.~Larson, S.~Mattu, and L.~Kirchner (2016).
\newblock Machine bias: There’s software used across the country to predict
  future criminals. and it’s biased against blacks.
\newblock {\em ProPublica\/}.

\bibitem[\protect\citeauthoryear{Berk, Heidari, Jabbari, Kearns, and Roth}{Berk
  et~al.}{2017}]{berk2017fairness}
Berk, R., H.~Heidari, S.~Jabbari, M.~Kearns, and A.~Roth (2017).
\newblock Fairness in criminal justice risk assessments: the state of the art.
\newblock {\em arXiv preprint arXiv:1703.09207\/}.

\bibitem[\protect\citeauthoryear{Byar and Green}{Byar and
  Green}{1980}]{ByarGreen1980}
Byar, D.~P. and S.~B. Green (1980).
\newblock The choice of treatment for cancer patients based on covariate
  information: Application to prostate cancer.
\newblock {\em Bulletin du Cancer\/}~{\em 67\/}(4), 477--490.

\bibitem[\protect\citeauthoryear{Chouldechova}{Chouldechova}{2017}]{chouldechova2017fair}
Chouldechova, A. (2017).
\newblock Fair prediction with disparate impact: A study of bias in recidivism
  prediction instruments.
\newblock {\em Big data\/}~{\em 5\/}(2), 153--163.

\bibitem[\protect\citeauthoryear{Cramer}{Cramer}{1946}]{Cramer46}
Cramer, H. (1946).
\newblock {\em Mathematical Methods of Statistics}.
\newblock Princeton University Press.

\bibitem[\protect\citeauthoryear{Dieterich, Mendoza, and Brennan}{Dieterich
  et~al.}{2016}]{dieterich2016compas}
Dieterich, W., C.~Mendoza, and T.~Brennan (2016).
\newblock Compas risk scales: Demonstrating accuracy equity and predictive
  parity.
\newblock {\em Northpoint Inc\/}.

\bibitem[\protect\citeauthoryear{Fan, Liu, Ning, and Zou}{Fan
  et~al.}{2017}]{Fanetal2017}
Fan, J., H.~Liu, Y.~Ning, and H.~Zou (2017).
\newblock High dimensional semiparametric latent graphical model for mixed
  data.
\newblock {\em Journal of the Royal Statistical Society: Series B (Statistical
  Methodology)\/}~{\em 79\/}(2), 405--421.

\bibitem[\protect\citeauthoryear{Friedman, Hastie, and Tibshirani}{Friedman
  et~al.}{2008}]{Friedman08}
Friedman, J., T.~Hastie, and R.~Tibshirani (2008).
\newblock Sparse inverse covariance estimation with the graphical lasso.
\newblock {\em Biostatistics\/}~{\em 9\/}(3), 432--441.

\bibitem[\protect\citeauthoryear{Goodman and Kruskal}{Goodman and
  Kruskal}{1954}]{goodman}
Goodman, L.~A. and W.~H. Kruskal (1954).
\newblock Measures of association for cross classifications.
\newblock {\em Journal of the American Statistical Association\/}~{\em
  49\/}(268), 732--764.

\bibitem[\protect\citeauthoryear{Hunt and Jorgensen}{Hunt and
  Jorgensen}{1999}]{Hunt1999}
Hunt, L. and M.~Jorgensen (1999).
\newblock Mixture model clustering using the multimix program.
\newblock {\em Australian \& New Zealand Journal of Statistics\/}~{\em
  41\/}(2), 154--171.

\bibitem[\protect\citeauthoryear{Johndrow and Lum}{Johndrow and
  Lum}{2017}]{johndrow2017algorithm}
Johndrow, J.~E. and K.~Lum (2017).
\newblock An algorithm for removing sensitive information: application to
  race-independent recidivism prediction.
\newblock {\em arXiv preprint arXiv:1703.04957\/}.

\bibitem[\protect\citeauthoryear{Kendall}{Kendall}{1948}]{Kendall1948}
Kendall, M. (1948).
\newblock {\em Rank Correlation Methods}.
\newblock C. Griffin.

\bibitem[\protect\citeauthoryear{Kendall}{Kendall}{1938}]{kendall1938}
Kendall, M.~G. (1938).
\newblock A new measure of rank correlation.
\newblock {\em Biometrika\/}~{\em 30\/}(1/2), 81.

\bibitem[\protect\citeauthoryear{Kendall}{Kendall}{1945}]{kendall1945}
Kendall, M.~G. (1945).
\newblock The treatment of ties in ranking problems.
\newblock {\em Biometrika\/}~{\em 33\/}(3), 239--251.

\bibitem[\protect\citeauthoryear{Kleinberg, Mullainathan, and
  Raghavan}{Kleinberg et~al.}{2016}]{kleinberg2016inherent}
Kleinberg, J., S.~Mullainathan, and M.~Raghavan (2016).
\newblock Inherent trade-offs in the fair determination of risk scores.
\newblock {\em arXiv preprint arXiv:1609.05807\/}.

\bibitem[\protect\citeauthoryear{Larson, Mattu, Kirchner, and Angwin}{Larson
  et~al.}{2016}]{larson2016we}
Larson, J., S.~Mattu, L.~Kirchner, and J.~Angwin (2016).
\newblock How we analyzed the compas recidivism algorithm.
\newblock {\em ProPublica (2016)\/}~{\em 9}.

\bibitem[\protect\citeauthoryear{Liu, Han, Yuan, Lafferty, and Wasserman}{Liu
  et~al.}{2012}]{Liu2012}
Liu, H., F.~Han, M.~Yuan, J.~Lafferty, and L.~Wasserman (2012).
\newblock High-dimensional semiparametric gaussian copula graphical models.
\newblock {\em The Annals of Statistics\/}~{\em 40\/}(4), 2293--2326.

\bibitem[\protect\citeauthoryear{Liu, Lafferty, and Wasserman}{Liu
  et~al.}{2009}]{Liu2009}
Liu, H., J.~Lafferty, and L.~Wasserman (2009).
\newblock The nonparanormal: Semiparametric estimation of high dimensional
  undirected graphs.
\newblock {\em Journal of Machine Learning Research\/}~{\em 10}, 2295--2328.

\bibitem[\protect\citeauthoryear{McParland and Gormley}{McParland and
  Gormley}{2016}]{McParland}
McParland, D. and I.~C. Gormley (2016).
\newblock Model based clustering for mixed data: clustmd.
\newblock {\em Advances in Data Analysis and Classification\/}~{\em 10\/}(2),
  155--169.

\bibitem[\protect\citeauthoryear{Rabe-Hesketh and Skrondal}{Rabe-Hesketh and
  Skrondal}{2007}]{SkrondalR2007}
Rabe-Hesketh, S. and A.~Skrondal (2007).
\newblock Latent variable modelling: A survey.
\newblock {\em Scandinavian Journal of Statistics\/}~{\em 34\/}(4), 712--745.

\bibitem[\protect\citeauthoryear{Somers}{Somers}{1962}]{somers}
Somers, R.~H. (1962).
\newblock A new asymmetric measure of association for ordinal variables.
\newblock {\em American Sociological Review\/}~{\em 27\/}(6), 799--811.

\bibitem[\protect\citeauthoryear{Tan, Caruana, Hooker, and Lou}{Tan
  et~al.}{2017}]{tan2017detecting}
Tan, S., R.~Caruana, G.~Hooker, and Y.~Lou (2017).
\newblock Detecting bias in black-box models using transparent model
  distillation.
\newblock {\em arXiv preprint arXiv:1710.06169\/}.

\bibitem[\protect\citeauthoryear{Tong}{Tong}{1990}]{tong}
Tong, Y. (1990).
\newblock {\em The Multivariate Normal Distribution}.
\newblock Springer, New York.

\bibitem[\protect\citeauthoryear{Xue and Zou}{Xue and Zou}{2012}]{XueZou2012}
Xue, L. and H.~Zou (2012).
\newblock Regularized rank-based estimation of high-dimensional nonparanormal
  graphical models.
\newblock {\em The Annals of Statistics\/}~{\em 40\/}(5), 2541--2571.

\bibitem[\protect\citeauthoryear{Zhou, Zhou, and Hooker}{Zhou
  et~al.}{2018}]{zhou2018approximation}
Zhou, Y., Z.~Zhou, and G.~Hooker (2018).
\newblock Approximation trees: Statistical stability in model distillation.
\newblock {\em arXiv preprint arXiv:1808.07573\/}.

\end{thebibliography}

\end{document}